\documentclass[manuscript]{acmart}
\usepackage{enumitem}

\AtBeginDocument{%
  }

\setcopyright{acmlicensed}
\copyrightyear{2025}
\acmYear{2025}
\acmDOI{XXXXXXX.XXXXXXX}

\acmJournal{JACM}
\acmVolume{37}
\acmNumber{4}
\acmArticle{111}
\acmMonth{4}

\usepackage[para]{footmisc}

\begin{document}

\title{SoK: Enhancing Privacy-Preserving Software Development from a Developers' Perspective}

\author{Tharaka Wijesundara}
\email{s4063322@student.rmit.edu.au}
\affiliation{%
  \institution{RMIT}
  \city{Melbourne}
  \state{Victoria}
  \country{Australia}
}

\author{Matthew Warren}
\affiliation{%
  \institution{RMIT}
  \city{Melbourne}
  \state{Victoria}
  \country{Australia}}
\email{matthew.warren2@rmit.edu.au}

\author{Nalin Arachchilage}
\affiliation{%
  \institution{RMIT}
  \city{Melbourne}
  \state{Victoria}
  \country{Australia}}
\email{nalin.arachchilage@rmit.edu.au}

\renewcommand{\shortauthors}{Trovato et al.}

\begin{abstract}
In software development, privacy preservation has become essential with the rise of privacy concerns and regulations such as GDPR and CCPA. While several tools, guidelines, methods, methodologies, and frameworks have been proposed to support developers embedding privacy into software applications, most of them are proofs-of-concept without empirical evaluations, making their practical applicability uncertain. These solutions should be evaluated for different types of scenarios (e.g., industry settings such as rapid software development environments, teams with different privacy knowledge, etc.) to determine what their limitations are in various industry settings and what changes are required to refine current solutions before putting them into industry and developing new developer-supporting approaches. For that, a thorough review of empirically evaluated current solutions will be very effective. However, the existing secondary studies that examine the available developer support provide broad overviews but do not specifically analyze empirically evaluated solutions and their limitations. Therefore, this Systematic Literature Review (SLR) aims to identify and analyze empirically validated solutions that are designed to help developers in privacy-preserving software development. The findings will provide valuable insights for researchers to improve current privacy-preserving solutions and for practitioners looking for effective and validated solutions to embed privacy into software development.
\end{abstract}





\maketitle

\section{Introduction} \label{introduction}

In the recent past, various tools, guidelines, methods, methodologies, and frameworks have been proposed, providing different types of support for developers embedding privacy into software applications, including privacy requirement specification, improving privacy awareness, privacy-preserving coding, etc. \cite{TangYing21, weirCharles23, PeixotoMarianaPCMTool, nguyen17, privacycat}. Despite the availability of these solutions, developers still face challenges when embedding privacy into software applications. For example, developers struggle to comply with regulations because of a lack of actionable technical guidelines \cite{TangYing21, EarpJulia, CompagnaLuca, ShapiroStuart10, weirCharles23}. Additionally, developers often lack privacy awareness, and due to that, they face difficulties in translating privacy principles into software requirements, which leads to deprioritizing privacy in software development \cite{PeixotoMarianaEvaluating, canedo22, BaldassarreMaria, AljeraisyAtheer24}. Even though tools, guidelines, methods, methodologies, and frameworks have been introduced, they come with their own limitations, such as requiring technical expertise to use, limited evaluation with developers, and needing manual intervention of developers \cite{AljeraisyAtheer24, PeixotoMarianaPCMTool, tianshi21}. 

Failing to properly embed privacy into software applications will negatively impact both users and organizations \cite{canedo22, oaic2019a, alzab21}. The Facebook-Cambridge Analytica scandal in 2018 is an example of how poor privacy management can result in major data misuse \cite{fbcam08}. It was about utilizing users’ data of 50 million Facebook profiles without their consent to better target political messages \cite{fbcam08}. It caused a drop in the users' trust in Facebook from 79\% to 28\% and resulted in a \$750 million lawsuit imposed by the European Union (EU) \cite{fbcam08}. This incident exemplifies the need for privacy-aware development practices to prevent data breaches, legal penalties, and reputational damage.

Recognizing these risks, governments established strict data protection rules such as General Data Protection Regulation (GDPR) \cite{gdpr16}, California Consumer Privacy Act (CCPA) \cite{ccpa18}, the Australian Privacy Act \cite{apa88}, and the New Zealand Privacy Act \cite{nzpa20} to ensure consistent privacy-preserving user data handling practices across organizations. However, simply having regulations in place does not guarantee compliance since developers lack the necessary guidance and technical support to implement them effectively \cite{AyalaRiveraGracePeriod, aljeraisy21}. 

To address these concerns, privacy-by-design (PbD) was introduced, allowing designers and developers to incorporate privacy practices into software systems from the early phase of software development and continuing it throughout the development cycle \cite{chassang2017impact, TangYing21, cavoukian09}. PbD is a core concept in GDPR and similar regulations \cite{chavesbenitti23}, implying that privacy and data protection measures should be integrated into systems from the beginning rather than as an afterthought. However, the lack of practical, developer-friendly guidelines, methods, methodologies, frameworks, and tools to translate its principles into a practical context and assist software developers in embedding privacy into software applications makes implementation difficult \cite{andrade23, trujillo19, lshammariMajed978, PedroBarbosa, chavesbenitti23}. 

These developer-facing challenges and the limitations of the existing solutions highlight the need for more effective, developer-centric privacy support mechanisms that can be seamlessly incorporated into existing software development practices. Therefore, to help developers embed privacy into software development by developing such support mechanisms, it is crucial to first have a broad understanding of the existing tools, guidelines, methods, methodologies, and frameworks. Additionally, it is important to understand the limitations of those proposed solutions in order to analyze how well those solutions addressed the developer challenges. To do that, we conduct a Systematic Literature Review (SLR) to find answers to these research questions:

\begin{enumerate}[label={\textbf{RQ\arabic*:}}, left=0pt]
\item What tools, guidelines, methods, methodologies, and frameworks are available as solutions to support software developers embedding privacy in software application development?

\item What are the limitations of the solutions identified in RQ1?

\end{enumerate}

The remaining sections of this paper are organized as follows. Section \ref{related-work} briefly presents an overview of existing research related to the field of this SLR, highlighting the gap that this SLR addresses. Section \ref{methodology} outlines the approach we used to identify, select, and analyze relevant literature. Then, in Section \ref{results}, the key findings from the selected studies are presented. In Section \ref{discussion}, by interpreting the results, we discuss how the solutions contribute to the field, whether they could address the developer challenges, and the suggestions for future research. Finally, Section \ref{theats_val}, \ref{limitations}, and \ref{conclusion} respectively address the threats to validity, constraints of the review, and the main insights with key takeaways. 

\section{Related Work} \label{related-work}

This section is intended to provide an overview of the current state of secondary studies, which specifically examined the existing developer support in privacy-preserving software development in the current literature. 

Trujillo et al. \cite{trujillo19} conducted a systematic mapping study to identify the available patterns, models, methods, and tools to support privacy-aware software development. The study highlighted that the majority of primary papers discussed privacy requirements and privacy design patterns, but with limited empirical validation \cite{trujillo19}. Chaves and Benitti later extended this study by reviewing papers published after 2018 to examine what has changed in the area \cite{chavesbenitti23}. They observed an increase in research that focused on embedding privacy throughout the development lifecycle \cite{chavesbenitti23}. However, they emphasized that these solutions have concerns about their practical applicability since they are yet to be validated in the industry settings \cite{chavesbenitti23}. 

Canedo et al. \cite{canedo22} studied a specific area of privacy-preserving software development, which is privacy requirements elicitation. The study identified various methodologies, techniques, and tools developed to help developers in gathering privacy requirements for software systems \cite{canedo22}. However, the study highlighted that these solutions lack empirical evaluation, leaving a concern about their effectiveness in practical scenarios \cite{canedo22}.

Overall, the results of the secondary studies highlight that most of the existing solutions that support developers in different stages of privacy-preserving software development are just proofs-of-concept \cite{canedo22, trujillo19, chavesbenitti23}. Also, most of them haven't been implemented in practical scenarios, and there is no empirical evidence \cite{canedo22, trujillo19, chavesbenitti23}. Current SLRs have analyzed those privacy-supporting solutions in a generalized manner, considering tools, guidelines, methods, methodologies, and frameworks collectively, without specifically focusing on empirically evaluated solutions. Furthermore, they do not examine the limitations of those proposed solutions and the limitations identified through empirical evaluation. 

To effectively improve existing solutions or propose new developer-supporting tools, guidelines, methods, methodologies, and frameworks, empirically evaluated results may provide valuable insights as they provide insights about practical usability, effectiveness, and the limitations faced by developers. Without having such an SLR, it may not be possible to analyze whether the proposed solutions address practical challenges in software development or are just theoretical contributions.

Therefore, this SLR takes a developer-centric approach, specifically focusing on empirically evaluated tools, guidelines, methods, methodologies, and frameworks that provide practical support for developers in privacy-preserving software development. Through this SLR, we aim to identify the problems in current solutions and provide insights to find gaps for further research.

\section{Methodology}
\label{methodology}

To find out how the literature has answered the research questions \textit{RQ1}, \textit{RQ2}, we conducted an SLR following the guidelines proposed by Kitchenham and Charters, which is a scientific and reproducible approach to conducting SLRs in software engineering \cite{kitchenham09}. The SLR spans across three stages: \textit{planning, conducting, and reporting}. This section covers the planning and conducting stages through different subsections, and the reporting section is covered in Section \ref{results}.

\subsection{Planning the SLR}
For the planning stage, we established a protocol that included four components: formulating research questions, developing search strings, selecting data sources, and selecting study criteria. The importance of the protocol in this stage was to reduce the researcher bias by establishing components before conducting the SLR \cite{kitchenham09}.  

\subsubsection{Formulating the research question}

The most important part of the SLR, as well as the planning stage, is formulating the research questions \cite{kitchenham09}. The research questions derive the entire SLR methodology by involving searching, extracting, and analyzing data \cite{kitchenham09}. In other words,
\begin{itemize}
 \item \textit{Data Searching}: The search strings should be related to the research questions \cite{kitchenham09}. 
 \item \textit{Data Extracting}: The extracted data should address the research questions \cite{kitchenham09}. 
 \item \textit{Data Analyzing}: The data should be analyzed in a way that can answer the research questions \cite{kitchenham09}.
\end{itemize}    

Therefore, we used a framework called PICOC (Population, Intervention, Comparison, Outcome, Context), which includes five criteria, to formulate research questions \cite{kitchenham09, petticrew06} as shown in Table \ref{table:picoc}. The "comparison" element was ignored in this SLR. Since one of the intentions of conducting this SLR was to find existing tools, guidelines, methods, methodologies, and frameworks as discussed in Section \ref{introduction}, there was nothing to compare with the "intervention". Further, the derived keyword(s) under each PICOC element, as shown in Table \ref{table:keywords} were used to develop search strings.

\begin{table*}
\footnotesize
  \caption{PICOC elements, their definitions, and SLR applications \cite{kitchenham09, petticrew06, fernandez18}. N/A - Not applicable}
  \label{table:picoc}
\begin{center}
\begin{tabular}
{|p{0.1\textwidth}|p{0.3\textwidth}|p{0.5\textwidth}|} 
 \hline
 \textbf{Element} & \textbf{Definition} & \textbf{SLR Application}\\ [0.5ex] 
 \hline\hline
 Population & What is the population that the SLR is interested in? & entities involved in privacy-preserving software development\\ 
 \hline
 Intervention & What are the existing approaches to address the core problem? & tools, guidelines, methods, methodologies, and frameworks \\
 \hline
 Comparison & What existing approaches can be compared with the intervention? & N/A\\
 \hline
 Outcome(s) & What are the results or effects of the interventions on the population? & helping to embed privacy in software development, resulting in privacy-aware software development\\
 \hline
 Context & What is the setting or environment in which the research is conducted? & software application development\\
 \hline
\end{tabular}
\end{center}
\end{table*}

\begin{table*}
\footnotesize
  \caption{Derivation of keywords for each PICOC element to formulate research questions and to build search strings. The "Comparison" element is ignored as it is not related to our SLR.}
  \label{table:keywords}
\begin{center}
\begin{tabular}
{|p{0.1\textwidth}|p{0.2\textwidth}|p{0.6\textwidth}|} 
 \hline
 \textbf{Element} & \textbf{Primary Phrase(s)} & \textbf{Derived Keyword(s)}\\ [0.5ex] 
 \hline\hline
 Population & software developers & developer*, "software engineer", "software engineers", programmer, "software designer", coder, "software practitioner", "software specialist"\\ 
 \hline 
 Intervention & available developer support to embed privacy & tool*, framework?, guideline?, pattern?, process, method*, technique?, model?, education, educational, knowledge, train*, intervention, awareness, practice?, support*, behaviour*, behavior* \\
 \hline
 Outcome(s) & privacy-embedded software, privacy awareness &  privacy \\
 \hline
 Context & software development & designing, coding, programming, verification, deployment, integration, "code review", validating, validation, testing\\
 \hline
\end{tabular}
\end{center}
\end{table*}

\subsubsection{Developing the Search Strings}
\label{section:develop_search_strings} 
We developed search strings to identify relevant literature using the derived keywords in Table \ref{table:keywords} as search terms. First, we derive primary phrases for each PICOC element based on its SLR application as listed under the column "Primary Phrase(s)" in Table \ref{table:keywords}. Then, we expanded the set of keywords by deriving keywords from each primary phrase (e.g., for software developers, we derived keywords such as software engineer, programmer, developer, and so on) and considering synonyms between the derived keywords (e.g., education - knowledge). The final list of derived keywords is listed under the column "Derived Keyword(s)" in Table \ref{table:keywords}.

Further, we used \textbf{asterisk wildcard (*)} for some terms to ensure that the relevant literature covered by different forms of the same word isn't ignored in the search process. For instance, we used "tool*" to cover  "tool", "tooling", "tooling", etc. We used \textbf{question marks (?)} to extend the possible keywords that are different from one character (e.g., framework? will take into account both framework and frameworks). And, to search with some exact phrases, we used \textbf{quotation marks ("")}. During this process, we make sure to remove the duplicate keywords found under each element. Finally, logical operators such as AND and OR were used to refine and advance the search query, which is as follows:

\textit{\small(developer* \textcolor{orange}{OR} "software engineer" \textcolor{orange}{OR} "software engineers" \textcolor{orange}{OR} programmer \textcolor{orange}{OR} "software designer" \textcolor{orange}{OR} coder \textcolor{orange}{OR} "software practitioner" \textcolor{orange}{OR} "software specialist") \textcolor{blue}{AND} (tool* \textcolor{orange}{OR} framework? \textcolor{orange}{OR} guideline? \textcolor{orange}{OR} pattern? \textcolor{orange}{OR} process \textcolor{orange}{OR} method* \textcolor{orange}{OR} technique? \textcolor{orange}{OR} model? \textcolor{orange}{OR} education \textcolor{orange}{OR} educational \textcolor{orange}{OR} knowledge \textcolor{orange}{OR} train* \textcolor{orange}{OR} intervention \textcolor{orange}{OR} awareness \textcolor{orange}{OR} practice? \textcolor{orange}{OR} support* \textcolor{orange}{OR} behaviour* \textcolor{orange}{OR} behavior*) \textcolor{blue}{AND} (privacy) \textcolor{blue}{AND} (software \textcolor{orange}{OR} designing \textcolor{orange}{OR} coding \textcolor{orange}{OR} programming \textcolor{orange}{OR} develop \textcolor{orange}{OR} development \textcolor{orange}{OR} verification \textcolor{orange}{OR} deployment \textcolor{orange}{OR} integration \textcolor{orange}{OR} "code review" \textcolor{orange}{OR} validating \textcolor{orange}{OR} validation \textcolor{orange}{OR} testing) \textcolor{blue}{AND} (software \textcolor{orange}{OR} code \textcolor{orange}{OR} program \textcolor{orange}{OR} algorithm \textcolor{orange}{OR} design \textcolor{orange}{OR} requirement \textcolor{orange}{OR} test \textcolor{orange}{OR} validate \textcolor{orange}{OR} review \textcolor{orange}{OR} debug \textcolor{orange}{OR} develop \textcolor{orange}{OR} development)}

\subsubsection{Selecting the Data Sources}
\label{section:selecting_data_source}

Once the search strings and the query were finalized, the next step was to identify the data sources from which relevant literature could be extracted. We considered three types of sources: digital libraries, journals, and conference proceedings. As digital libraries, we considered, IEEE Xplore\footnote{https://ieeexplore.ieee.org}, ACM Digital Library\footnote{https://dl.acm.org}, Scopus\footnote{https://www.scopus.com/}, SpringerLink\footnote{https://link.springer.com/}, and Google Scholar\footnote{https://scholar.google.com}. To select journals and conferences, first, we categorized the scope of the literature into four disciplines, considering the research questions. The disciplines include computer security, privacy and cryptography, software systems, artificial intelligence, and human-computer interaction (HCI). Then, we selected top-tier venues for each category, as shown in Table \ref{table:venues}. In total, we considered 12 conferences and 10 journal publications to conduct the SLR.

\begin{table*}
\scriptsize
\caption{Selected top-tier venues for each discipline}
\label{table:venues}
\begin{center}
\begin{tabular}{|p{0.10\textwidth}|p{0.43\textwidth}|p{0.43\textwidth}|} 
 \hline
 \textbf{Discipline} & \textbf{Conferences} & \textbf{Journals}\\ [0.5ex] 
 \hline\hline
 Computer Security and Cryptography &
        \vspace{-0.5\baselineskip}\begin{itemize}[label=-, leftmargin=*, nosep] 
            \item IEEE Symposium on Security and Privacy (S\&P)
            \item ACM Symposium on Computer and Communications Security (CCS)
            \item USENIX Security Symposium
            \item Network and Distributed System Security Symposium (NDSS)
            \item Symposium On Usable Privacy and Security (SOUPS)
            \item Proceedings on Privacy Enhancing Technologies Symposium (PoPETS)
            
        \vspace{-\baselineskip}\end{itemize} 

         & \vspace{-0.5\baselineskip}\begin{itemize}[label=-, leftmargin=*, nosep] 
            \item IEEE Transactions on Information Forensics and Security (TIFS)
            \item Computers \& Security (Com. \& Sec.)
            \item IEEE Transactions on Dependable and Secure Computing (TDSC)
            \item Security and Communication Networks (Sec. \& Com. Net.)
            \item ACM Transactions on Privacy and Security (TOPS)
        \vspace{-\baselineskip}\end{itemize}
        \\ 
 \hline
 Software Systems & 
        \vspace{-0.5\baselineskip}\begin{itemize}[label=-, leftmargin=*, nosep] 
            \item International Conference on Software Engineering  (ICSE)
            \item International Conference on Automated Software Engineering (ASE)
            \item ACM SIGSOFT International Symposium on Foundations of Software Engineering (SIGSOFT)
        \vspace{-\baselineskip}\end{itemize}  & 
        \vspace{-0.5\baselineskip}\begin{itemize}[label=-, leftmargin=*, nosep] 
            \item IEEE Transactions on Software Engineering (TSE)
            \item ACM Transactions on Software Engineering and Methodology (TOSEM)

        \vspace{-\baselineskip}\end{itemize}\\ 
 \hline

 Human-Computer Interaction (HCI) & 
        \vspace{-0.5\baselineskip}\begin{itemize}[label=-, leftmargin=*, nosep] 
            \item ACM Conference on Human Factors in Computing Systems (CHI) 
            \item Proceedings of the ACM on Human-Computer Interaction (HCI)
            \item Computer Supported Cooperative Work (CSCW)
        \vspace{-\baselineskip}\end{itemize} & 
        \vspace{-0.5\baselineskip}\begin{itemize}[label=-, leftmargin=*, nosep] 
            \item IEEE Transactions on Human-Machine Systems (Trans. Hum. Mach. Sys.)
            \item ACM Transactions on Computer-Human Interaction (TOCHI)
            \item IEEE Transactions on Mobile Computing (Trans. Mob. Com.)
        \vspace{-\baselineskip}\end{itemize}\\ 
 \hline
\end{tabular}
\end{center}
\end{table*}

\subsubsection{Study Selection Criteria}
Then, we defined the selection criteria to decide what studies should be included or excluded from the SLR \cite{kitchenham09} through Inclusion Criteria (IN) and Exclusion Criteria (EX) as shown in Table \ref{table:incexl}. Criteria were defined in a way that extracts the most relevant publications to address the research questions. To define the criteria, we gain insights from an SLR on privacy-enhancing technologies in software development \cite{boteju2023}.

\begin{table*}
\footnotesize
  \caption{Inclusion and Exclusion Criteria}
  \label{table:incexl}
\begin{center}
\begin{tabular}
{|p{1\textwidth}|} 
 \hline
 \textbf{Inclusion Criteria} \\
 \hline\hline
 \textit{IN1}: The publication should address at least one of the research questions. \\ 
 \hline
 \textit{IN2}: The publication should be published in a peer-reviewed journal or conference. \\ 
 \hline
 \textit{IN3}: The publication should include empirical evidence from software developers or students who have a software background. The proposed developer support should be used by the developers (e.g., there is a tool that generates privacy policy statements. The authors generated some statements and got some feedback from developers about the generated statements. However, since the tool has not been used by the developers, it will not be included in our SLR.). \\ 
 \hline
 \textit{IN4}: The entire publication or a portion of it should be privacy-focused. (e.g., publications may address both security and privacy together).\\
 \hline
 \textit{IN5}: The publication should address software developers who engage in software development. (e.g., we consider publications that address software developers who do both developing and designing).\\
 \hline
 \textit{IN6}: The publication should be written in the English language.\\ 
 \hline
 \textbf{Exclusion Criteria} \\
 \hline\hline
 \textit{EX1}: The publication focuses on applications outside the software development unless it provides insights about addressing \textit{RQs} of this SLR.\\ 
 \hline
 \textit{EX2}: The publication is a short paper, poster, abstract, tutorial, extended abstract, demonstration, introduction, keynote, opinion paper, concept paper, editorial, work in progress, interview, monograph, secondary study, tertiary study, technical paper, book chapter, ongoing study, white paper.\\
\hline
\end{tabular}
\end{center}
\end{table*}

\subsection{Conducting the SLR}

Once the protocol was finalized, the next step was conducting the SLR. We conducted the SLR through three stages: data search and selection, data extraction, and data analysis \cite{kitchenham09}. 

\subsubsection{Study Search and Selection}

\begin{figure}[h]
  \centering
  \includegraphics[width=\linewidth]{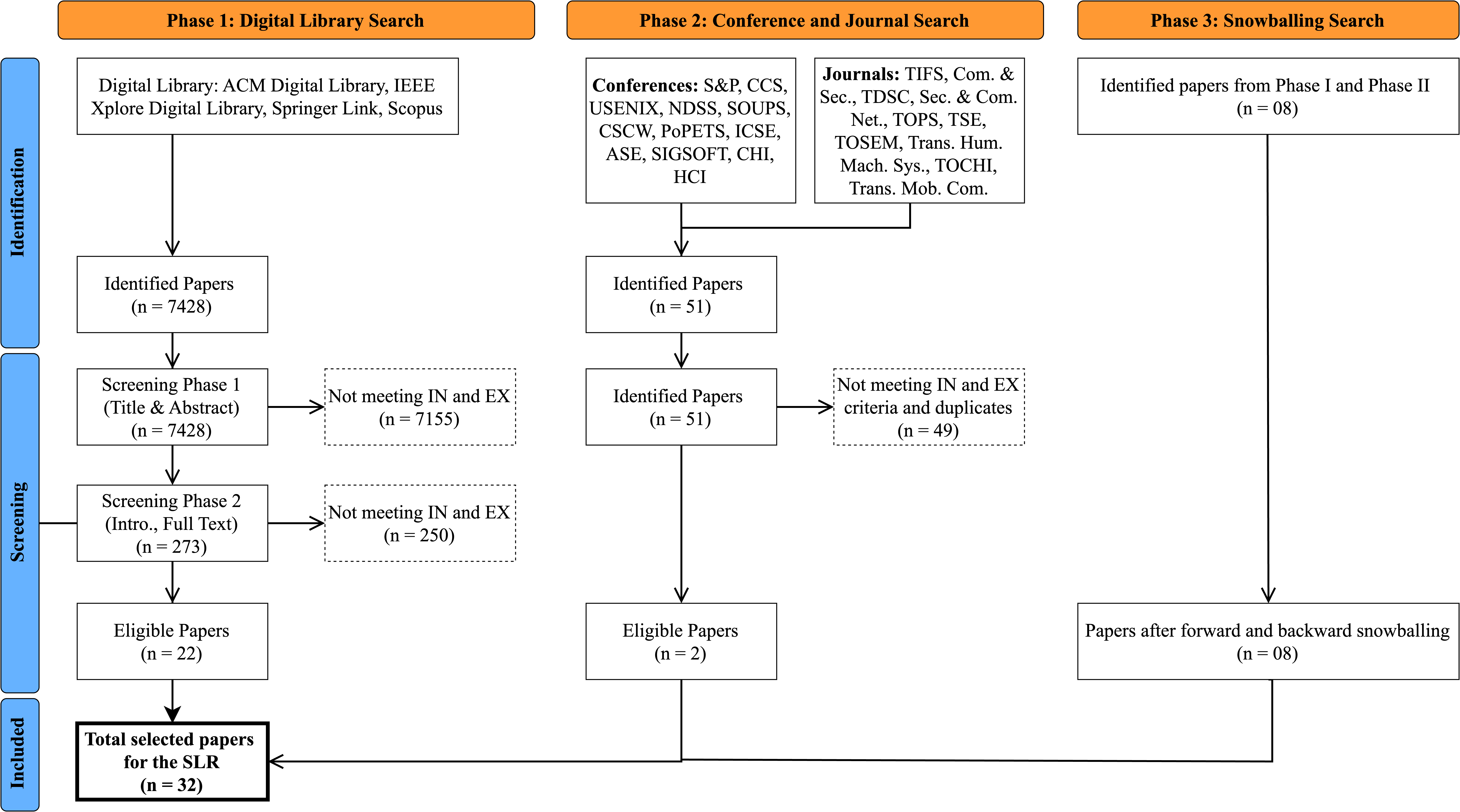}
  \caption{The summary of the study search and selection process using the cPRISMA framework. n - number of papers, IN - Inclusion Criteria, EX - Exclusion Criteria.}
  \Description{PRISMA paper selection graph}
  \label{prisma}
\end{figure}

The identification of relevant publications from data sources discussed in Section \ref{section:selecting_data_source} was conducted through three different phases. The study selection was performed in September 2024-January 2025 following three main phases: automatic search on digital libraries \cite{kitchenham09}, manual search on selected journals and conference proceedings \cite{kitchenham09}, and bidirectional snowballing \cite{snowball14}, as shown in Figure \ref{prisma}. 

For phases I and II, the search process was conducted through the data sources discussed in Section \ref{section:selecting_data_source} using a search query constructed in Section \ref{section:develop_search_strings}. During Phase III, we reviewed the references and citations of the studies identified in Phases I and II to ensure that all the relevant literature was considered for the SLR. Further, we didn't consider any time limitations for our search. Also, to manage publications during all three phases, we used Zotero Reference Management Software \footnote{https://www.zotero.org/}.

The constructed search query resulted in 7428 publications in phase I. Phase I was conducted in two stages. First, the titles and abstracts of all the resulting publications were reviewed while checking their suitability with inclusion and exclusion criteria. Then, 273 publications were selected for further review. Second, the review was extended towards the introduction, conclusion, and sometimes for the full text of the selected publications in the first step. After the second step, 22 eligible publications were selected for the SLR.

In phase II, the same search query was used at some venues where the search-by-keyword facility was incorporated (e.g., ACM and IEEE venues listed in Table \ref{table:venues} have that facility). Sometimes, we had to do a manual search through all the papers published in different years to select relevant papers (e.g., NDSS, PoPETS, etc.). During the search, we ignored the already selected papers in Phase I. After applying inclusion and exclusion criteria, we identified 02 suitable publications for the SLR. 

Once we finished Phases I and II, we conducted the snowballing technique as Phase III of our search based on the finalized papers from Phases I and II. We ignored some previously identified papers during Phase III, just as we did in Phase II. After all, 08 papers were selected.

Once all three phases were conducted, altogether 32 final publications were extracted to conduct the SLR. The overall study search and selection strategy is depicted in Figure \ref{prisma}. 

\subsubsection{Data Extraction and Analysis}
Once the relevant papers were selected, we extracted some data from each paper and tabularized them in a spreadsheet to familiarize ourselves with the content and the domain of focus of each paper. The table includes the title, authors, publication year, venue, purpose/objective of the paper, contribution, main findings, limitations, and future work. 

After the data extraction and familiarization step, we conducted the reflexive thematic analysis \cite{coding2}. We started coding the selected papers without using a pre-defined coding scheme, but with the research questions in mind \cite{coding2}. This allowed us to code papers based on a list of codes derived from the research questions and the authors' knowledge of privacy-preserving software development. Then, we expanded the list of codes by adding new codes based on the extracted data \cite{coding2}. For instance, after going through the extracted data multiple interactions, the phrase "make use of personal data while simultaneously making it easier to analyze how that personal data is processed....." \cite{privacystreams} was coded as "personal data", "tool", and "process" by the first author. 

Once finished coding, they were grouped under different themes based on the patterns and similarities of the identified codes in order to make use of them to answer the research questions, RQ1 and RQ2. For example, the publication with the above phrase was categorized under \textit{Manage \& Secure Personal Data}. Further, we considered the paper's main contribution when assigning them to the themes. For example, Li et al. \cite{privacystreams} support managing personal data in privacy-preserving application development while helping developers enhance their privacy education. However, since its main contribution is to provide developer support to manage personal data, we assigned it under the theme \textit{Manage \& Secure Personal Data}. Finally, we defined six different themes to answer the research questions, RQ1 and RQ2.

The co-authors validated the generated codes, generated themes' relevance to the research questions RQ1 and RQ2, as well as assigning publication under each theme. This was accomplished through multiple discussions among the co-authors to determine whether there were any inconsistencies in the coding, theming, or publication assignment process. To accomplish this, the first author created a spreadsheet containing the relevant details of the publications, the coding and theming details, and the assignment of publications to the generated themes. Then, the spreadsheet was discussed during the discussions to check for inconsistencies. During the discussions, the co-authors' diverse perspectives were considered, and thematic development and publication assignments were carried out according to the guidelines proposed by Braun and Clark \cite{Clarke2014}. 

\begin{figure}[h]
  \centering
  \includegraphics[width=\linewidth]{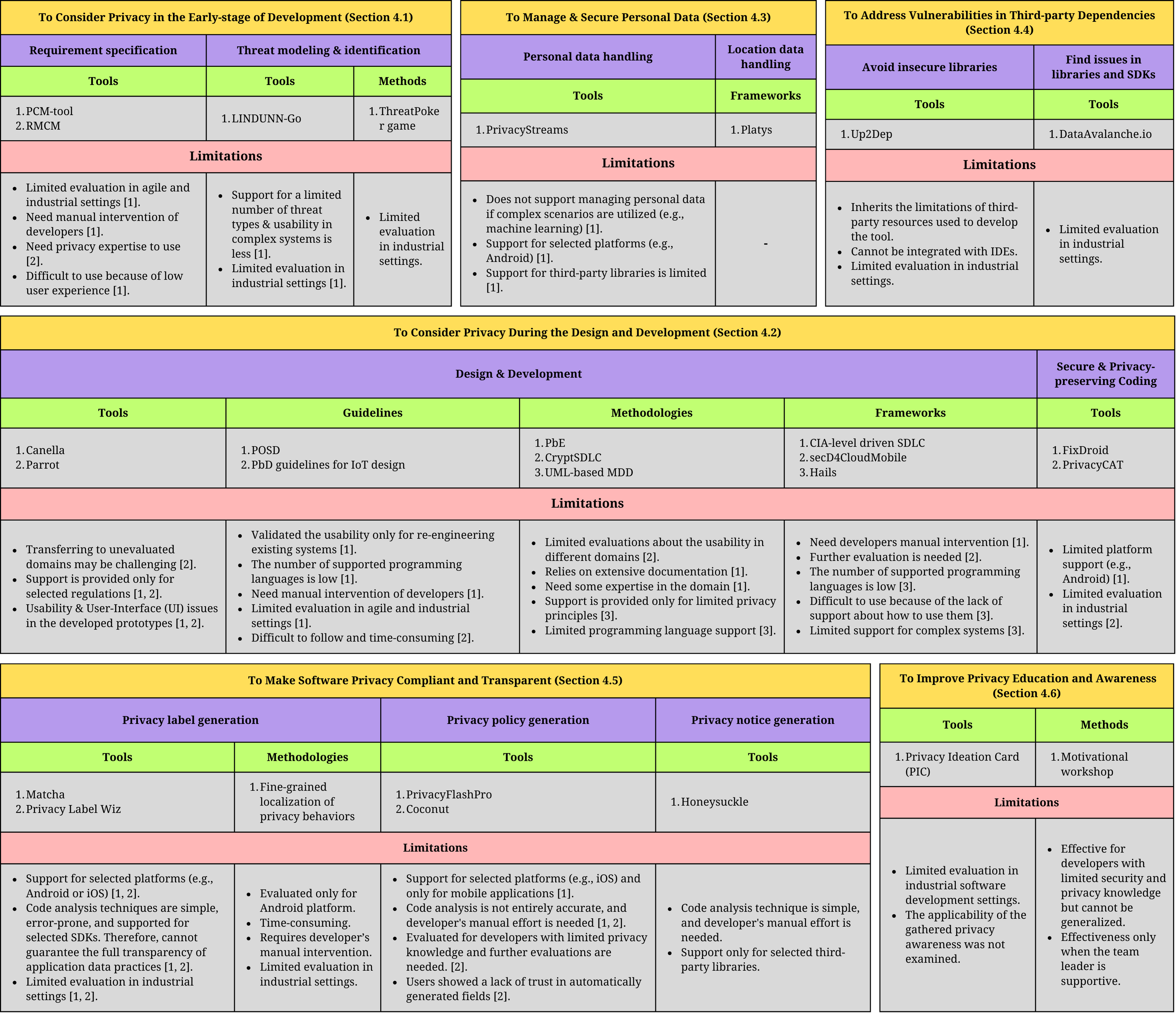}
  \caption{The summary of the thematic analysis. In RQ1, we discussed the available tools, guidelines, methods, methodologies, and frameworks as solutions. In RQ2, we discuss their limitations.}
  \label{summary_graph}
\end{figure}

\begin{table*}
\scriptsize
\caption{A summary of empirical evaluation contexts. Here, student refers to those who study at the university or school level, and industry refers to developers with industry experience (i.e., experience in industry-level projects).}
\label{table:evaluation}

\begin{minipage}{0.48\textwidth}
\begin{center}
\begin{tabular}{|p{0.20\textwidth}|p{0.12\textwidth}|p{0.35\textwidth}|} 
 \hline
 \textbf{Name} & \textbf{Reference} & \textbf{Evaluation context}\\ [0.5ex] 
 \hline
 PCM-tool & \cite{PeixotoMarianaEvaluating, PeixotoMariananatural, PeixotoMarianaPCMTool} & Students and industry \\ 
 \hline
 RMCM & \cite{MAI2018165} & Industry  \\ 
 \hline
 LINDUNN-Go & \cite{WuytsKim20} & Students and industry \\ 
 \hline
 ThreatPoker game & \cite{RyggeHanne} & Students \\ 
 \hline
 PrivacyStreams & \cite{privacystreams} & Students and industry \\ 
 \hline
 Platys & \cite{MurukannaiahPradeep} & Students \\ 
 \hline
 Up2Dep & \cite{nguyen20} & Students and industry \\ 
 \hline
 DataAvalanche.io & \cite{ekambaranathan23} & Industry \\ 
 \hline
 Canella & \cite{AljeraisyAtheer24} & Students and industry \\ 
 \hline
 Parrot & \cite{AlhirabiNada23, AlhirabiNada24} & Students and industry \\ 
 \hline
 POSD & \cite{BaldassarreMaria, BaldassarreTeresapkb} & Industry \\ 
 \hline
 PbD guidelines for IoT design & \cite{PERERA2020238} & Students \\ 
 \hline
 PbE & \cite{PedroBarbosa} & Students \\ 
 \hline
\end{tabular}
\end{center}
\end{minipage}
\hfill
\begin{minipage}{0.48\textwidth}
\begin{center}
\begin{tabular}{|p{0.30\textwidth}|p{0.12\textwidth}|p{0.35\textwidth}|} 
 \hline
 \textbf{Name} & \textbf{Reference} & \textbf{Evaluation context}\\ [0.5ex] 
 \hline
 CryptSDLC & \cite{LoruenserThomas18} & Industry \\ 
 \hline
 UML-based MDD & \cite{krstic24} & Did not clearly mention \\ 
 \hline
 CIA-level driven SDLC & \cite{KangSooyoung} & Industry \\ 
 \hline
 secD4CloudMobile & \cite{ChimucoFrancisco} & Did not clearly mention \\ 
 \hline
 Hails & \cite{GiffinDaniel12} & Did not clearly mention \\ 
 \hline
 FixDriod & \cite{nguyen17} & Students and industry \\ 
 \hline
 PrivacyCAT & \cite{privacycat} & Industry \\ 
 \hline
 Matcha & \cite{tianshi24} & Students and industry \\ 
 \hline
 Privacy Label Wiz & \cite{gardner22} & Did not clearly mention \\ 
 \hline
 Fine-grained localization of privacy behaviors & \cite{jain23} & Students and industry \\ 
 \hline
 PrivacyFlashPro & \cite{zimmeck2021PrivacyFlashPA} & Industry \\ 
 \hline
 Coconut & \cite{coconut} & Students and industry \\ 
 \hline
 Honeysuckle & \cite{tianshi21} & Industry \\ 
 \hline
 Privacy Ideation Card (PIC) & \cite{TangYing21} & Students \\ 
 \hline
 Moticational Workshop & \cite{weirCharles23} & Industry \\ 
 \hline
\end{tabular}
\end{center}
\end{minipage}

\end{table*}

\section{Results} \label{results}

As shown in Figure \ref{summary_graph}, this SLR presents an in-depth analysis of existing publications about developer-supporting guidelines, tools, methods, methodologies, and frameworks for incorporating privacy into software applications. Even though different types of solutions exist, developers struggle with issues such as not having good privacy knowledge to use them, not having proper solutions to be used in the Agile method, and some accuracy issues of the solutions \cite{PeixotoMarianaPCMTool, PedroBarbosa, tianshi21, coconut, zimmeck2021PrivacyFlashPA}. 

The results section delves into these issues by discussing the existing developer-supporting mechanisms to address these issues. As shown in Figure \ref{summary_graph}, to answer the research questions (RQ1 and RQ2), we categorized the identified developer-supporting mechanisms into six different themes (as depicted in yellow color) and sub-themes (as depicted in purple color) based on the type of support they provide for developers. For example, in Section \label{subsec:privacy-requirement}, we discuss the available developer-supporting mechanisms under the theme "consider privacy in the early stage of development" and two sub-themes: "requirement specification" and "threat modeling and identification.". However, please note that Section \ref{subsec:privacy-education} does not have any subthemes because the identified mechanisms were discussed under the main theme. 

Further, as shown in Figure \ref{table:evaluation}, we summarized the evaluation context of all the identified developer-supporting solutions, which we discuss further in both the results and discussion sections.

The following sections will mainly discuss the Figure \ref{summary_graph} in detail and also Figure \ref{table:evaluation}, answering RQ1, while RQ2 was answered in the discussion section (i.e., Section \ref{discussion}).

\subsection{Consider Privacy in the Early-stage of Development}
\label{subsec:privacy-requirement} 

\subsubsection{Requirement specification.}
Privacy is one of the major Non-Functional Requirements (NFRs) in Software Engineering (SE), which is often overlooked or given minimal attention during software development \cite{PeixotoMariananatural}. Addressing privacy requirements early in the development process may help avoid privacy violations that may happen later in the development process \cite{GharibMohamad20}. However, Agile software development (ASD) does not have a specific technique to guide developers to privacy requirement specifications in a situation where outdated privacy requirements pose challenges for agile teams \cite{PeixotoMarianaEvaluating, canedo22}. For example, assume an agile team is creating a healthcare app with old privacy rules, like collecting user health data without clear consent. When new laws, such as GDPR, require stricter data protection, the team finds it hard to update their app quickly.

To address this, Peixoto et al. \cite{PeixotoMarianaPCMTool} developed the Privacy Criteria Method tool (PCM-tool), which provides structured guidance for privacy requirement specification in Agile. The PCM tool has a user interface with input fields to specify each privacy requirement \cite{PeixotoMarianaPCMTool}. It consists of input fields such as a detailed description of privacy requirements, their priority level, and associated vulnerabilities, legal compliance (e.g., GDPR), ensuring that privacy requirements are updated and aligned with current regulations \cite{PeixotoMarianaPCMTool}. Further, some developer guides are available about how to fill these fields \cite{PeixotoMarianaPCMTool}. Finally, empirical evaluations with graduate students and industry professionals indicate that the PCM tool aids developers in specifying good-quality privacy requirements \cite{PeixotoMarianaEvaluating, PeixotoMariananatural, PeixotoMarianaPCMTool}. 

Separately, Mai et al. \cite{MAI2018165} proposed a method called Restricted Misuse Case Modeling (RMCM) that utilizes use cases and misuse cases to model security and privacy requirements. Here, a use case refers to an interaction between a user and the system to achieve a specific goal, while a misuse case refers to a scenario in which an attacker tries to compromise the security and privacy of the system. RMCM provides structured templates and restriction rules (i.e., constraints or conditions that limit how a system, process, or model should behave to ensure security or compliance) to clearly define security threats, threat scenarios, and mitigation strategies \cite{MAI2018165}. It ultimately helps to model privacy requirements \cite{MAI2018165}. The approach was successfully applied to an industrial healthcare project, demonstrating its effectiveness \cite{MAI2018165}.

\subsubsection{Threat modeling \& identification.}
Threat modeling is another essential technique for early-stage privacy risk assessment \cite{WuytsKim20}. However, current threat modeling methods like LINDUNN are too complex and require a lot of expertise and time for developers to understand \cite{WuytsKim20}. As a solution, Wuyts et al. \cite{WuytsKim20} proposed a simplified card-based toolkit for collaborative threat identification called LINDUNN-Go. LINDUNN-Go is a teamwork that allows people with different skills (developers, privacy experts, etc.) to work together for the privacy threat analysis \cite{WuytsKim20}. LINDUNN-Go consists of threat-type cards that describe potential privacy threats with examples and guidance in a way that can even be understood by non-experts \cite{WuytsKim20}. Each card follows the same template and includes necessary information such as threat source, consequences, the area where the threat occurs, etc \cite{WuytsKim20}. The feedback gathered from students and industry experts showed that LINDUNN-Go is easier to use and understand \cite{WuytsKim20}. 

Further, to identify threats in the software systems and the effort needed to mitigate vulnerabilities related to those threats during Agile development, Rygge and Jøsang \cite{RyggeHanne} introduced ThreatPoker, a team-based card game \cite{RyggeHanne}. The game consists of two rounds \cite{RyggeHanne}. First is the risk round, where team members discuss and analyze potential security and privacy threats associated with the user stories or features being developed in the current sprint \cite{RyggeHanne}. Also, during the same round, they estimated the severity of security and privacy risks by playing cards \cite{RyggeHanne}. Second is the solution round, where they estimate the effort required to address the threats \cite{RyggeHanne}. Since threat poker is a team game where everyone, including developers, shares their ideas and helps to make security and privacy-related decisions, it helps everyone feel responsible for keeping the project safe \cite{RyggeHanne}.

\subsection{Consider Privacy During the Design and Development Stages}
\label{subsec:privacy-integrate}

Ensuring privacy throughout the software design and development process is critical to mitigating the privacy-related problems that might arise later \cite{LoruenserThomas18, McGrawGary}. For example, if a healthcare app fails to encrypt sensitive patient data during development, hackers may later exploit this vulnerability and steal the data. In this section, we discuss the existing tools \cite{AlhirabiNada23, AljeraisyAtheer24, privacycat, nguyen17}, guidelines \cite{BaldassarreMaria, PERERA2020238}, methodologies \cite{LoruenserThomas18, PedroBarbosa, krstic24}, and frameworks \cite{KangSooyoung, ChimucoFrancisco, GiffinDaniel12} to design and develop applications in a privacy-preserving way. 

\subsubsection{Tool-based design and development}
\label{subsubsec:privacy-integrate-tooling}

IoT application development is more complicated than desktop, mobile, and web application development because of the collaborative work needed with different development backgrounds, such as software, hardware, and embedded systems \cite{PERERA2020238}. Parrot \cite{AlhirabiNada23} and Canella \cite{AljeraisyAtheer24} are tools specifically designed to help developers in privacy-aware IoT application design and development.

Incorporating privacy features into IoT is challenging for developers because the existing guidelines are difficult to translate into technical implementation due to reasons such as they are written in legal language \cite{AlhirabiNada23, TahaeiMohammad21, SenarathAwanthika18}. For example, terms like "lawful basis for processing" or "data subject rights" from GDPR \cite{gdpr16} require legal expertise to fully understand and implement. As a solution, Parrot comes as a developer-friendly tool with an interactive design interface that translates complex legal privacy requirements into actionable technical design choices \cite{AlhirabiNada23}. For example, PARROT simplifies GDPR concepts like "lawful basis for processing" or "data subject rights" by breaking them down into clear, configurable options and visually highlighting potential privacy risks while providing instant feedback \cite{AlhirabiNada23}. At the same time, it supports visualizing the system's architecture and data flows \cite{AlhirabiNada23}. Finally, it showed that PARROT can help novice IoT developers comply with privacy laws and follow privacy guidelines \cite{AlhirabiNada24}.

Separately, Canella supports privacy-aware IoT application development using visual programming tools such as Blockly and Node-RED, which support drag-and-drop-based development \cite{AljeraisyAtheer24}. It provides real-time feedback on potential privacy issues in the system while highlighting areas where user data might be at risk and suggesting less risky alternatives for handling personal information \cite{AljeraisyAtheer24}. Also, it includes a Privacy Law Validator to ensure that the IoT system complies with the data minimization principle as well as the privacy and data protection rules of the UK \cite{AljeraisyAtheer24}. Further, it offers interactive features like warning messages and tooltips to reduce the cognitive load on developers and enhance their understanding of privacy practices. Overall, it helps to protect user data and keep IoT applications up to date with evolving legal requirements \cite{AljeraisyAtheer24, hadar18, balebako14, LiTianshiLouie}.

\subsubsection{Guideline-based design and development}
\label{subsubsec:privacy-integrate-guideline}
Privacy guidelines are a set of clear instructions and best practices for developers to follow when embedding privacy into software. It is challenging for developers to translate these guidelines into technical implementations when they do not have the required expertise in privacy \cite{BaldassarreMaria}. For example, "Minimize the collection of personal data to only what is necessary for the intended purpose" is a guideline based on the data minimization principle of Privacy by Design \cite{cavoukian2012operationalizing} as well as GDPR \cite{gdpr16}. Privacy-Oriented Software Development (POSD) \cite{BaldassarreMaria} helps developers translate these privacy guidelines into technical implementations by providing developers with a structured, practical, and actionable technical implementation via a Privacy Knowledge Base (PKB) \cite{BaldassarreTeresapkb}. The PKB acts as a bridge between high-level privacy principles (e.g., Privacy by Design) and practical, actionable, technical implementations \cite{BaldassarreTeresapkb}. POSD's guidelines proposed a static code analyzer to identify vulnerabilities in the code \cite{BaldassarreMaria}. PKB helps map these identified vulnerabilities with privacy patterns, which are reusable solutions for common privacy problems (e.g., logging user activities that contain sensitive information for debugging purposes), to mitigate the vulnerabilities \cite{BaldassarreMaria}. These patterns include detailed descriptions, diagrams, and implementation steps, making it easier for developers to apply them in code \cite{BaldassarreMaria}.

Another set of guidelines was proposed by Perera et al. \cite{PERERA2020238} to assist developers in following the PbD principles when designing IoT applications. This was proposed to support embedding privacy specifically into heterogeneous and distributed IoT systems \cite{PERERA2020238}. This offers systematic guidance through a set of privacy guidelines and a method for applying them during the IoT application design process \cite{PERERA2020238}. It consists of 21 privacy guidelines derived from Hoepman’s privacy design strategies, which cover various aspects of privacy, such as data minimization, data anonymization, encryption, etc. \cite{HoepmanJaapHenk, PereraCharithMcCormickpbd16}.

\subsubsection{Methodology-based design and development}
\label{subsubsec:privacy-integrate-method}
As a solution to the lack of guidance provided by PbD to implement them in practice, Barbosa et al. \cite{PedroBarbosa} proposed a methodology called Privacy-by-Evidence (PbE) \cite{PedroBarbosa}. The methodology consists of a clear, step-by-step workflow that can be run in parallel to the regular development cycle \cite{PedroBarbosa}. The workflow consists of activities such as identifying the application context and data formats, checking compliance with regulations, assessing risks, applying privacy techniques (e.g., anonymization, generalization, noise addition), and evaluating potential attacks \cite{PedroBarbosa}. Here, evidence refers to documented proofs that explain how privacy risks have been identified, mitigated, and addressed throughout the software development process. PbE emphasizes documenting this evidence using Goal Structuring Notation (GSN) \cite{PedroBarbosa}. 

Even though PbE focuses on systematically incorporating privacy into software systems \cite{PedroBarbosa}, it does not address the technical complexities of cryptographic security, which requires expertise in the cryptographic engineering domain \cite{LoruenserThomas18}. For example, k-anonymity is a privacy-preserving technique for anonymization that requires expertise in cryptographic engineering to properly understand and implement its underlying mechanisms. Cryptographic protection is a critical component in privacy-preserving software development, which introduces additional vulnerabilities if implemented incorrectly \cite{LoruenserThomas18}. For example, improperly applying k-anonymity will lead to re-identification attacks. As a solution, Loruenser et al. \cite{LoruenserThomas18} proposed CryptSDLC, a methodological approach to ensure secure-by-design software systems by providing software developers with access to advanced cryptographic solutions and minimizing potential errors when integrating cryptographic functionality into software. It was achieved by offering cryptographic solutions in a simplified form by abstracting their complex cryptographic concepts, which eventually reduces the need for deep cryptographic expertise to use them in software systems and minimize integration errors \cite{LoruenserThomas18}. CryptSDLC facilitates communication and collaboration among different stakeholders, such as cryptographers, security experts, and developers, and bridges the communication gaps that often hinder the effective integration of cryptography into software development \cite{LoruenserThomas18}. This was achieved by providing communication tools such as Service Level Agreements (SLAs) and cryptographic design patterns to share security requirements, capabilities, and constraints with stakeholders.

Separately, Krstic et al. \cite{krstic24} proposed a Model-Driven Development (MDD) methodology that provides comprehensive support for developers in implementing privacy policies, particularly focusing on purpose limitation and consent as required by regulations like GDPR \cite{krstic24}. Developers work with three models: data model, security model, and privacy model \cite{krstic24}. These models are presented using Unified Modeling Language (UML) diagrams \cite{krstic24}. First, software engineers use the data model to describe how the system’s data is organized \cite{krstic24}. Next, security engineers use the security model to specify who can access or change the data \cite{krstic24}. Finally, privacy engineers use the privacy model to ensure that personal data is only used for specific purposes and only with the user’s consent \cite{krstic24}. Software engineers, security engineers, and privacy engineers collaboratively work to provide these models \cite{krstic24}. Then, these UML models are translated into actual system implementation codes in C\# (ASP.NET) and Python (Flask) \cite{krstic24}. It reduces the manual coding effort and minimizes human errors \cite{krstic24}. 

\subsubsection{Framework-based design and development}
\label{subsubsec:privacy-integrate-framework}

Security-by-design is a proactive approach that considers security during the early phase of software development \cite{KangSooyoung, avizienis}. Secure software development lifecycle (Secure-SDLC) refers to the development process that incorporates security-by-design \cite{KangSooyoung, avizienis}. Kang et al. \cite{KangSooyoung} proposed a CIA-level driven secure SDLC framework allowing organizations to tailor security practices according to the level of security they require and offer a structured and evidence-based approach to integrating security into the software development lifecycle \cite{KangSooyoung}. It maps 66 security activities across 10 phases of the SDLC, ensuring security is embedded at every stage, from the beginning to the end \cite{KangSooyoung}. For example, it consists of the "Security Requirements Elicitation" activity with the required guidance to implement it \cite{KangSooyoung}. Developers have been provided with detailed security activities, evidence templates, standards that the activity aims to achieve (e.g., ISO/IEC 27001), and tools to document and prove compliance with security requirements \cite{KangSooyoung}. The mapping helps developers to get a clear picture of the activities that must be completed at each phase to meet security standards. Further, it allows developers to compare their security practices with competitors and identify areas for improvement \cite{KangSooyoung}. 

Similarly, to specifically address security issues of cloud-based mobile applications through security-by-design, Chimuco et al. proposed a framework called secD4CloudMobile, which includes developer support covering different areas such as security requirement elicitation, attack modeling, and security testing \cite{ChimucoFrancisco}. These sets of solutions facilitate the incorporation of security mechanisms throughout the software development lifecycle \cite{ChimucoFrancisco}. It emphasizes incorporating security-by-design practices that indirectly protect sensitive user data and maintain privacy \cite{ChimucoFrancisco}. For example, helping developers to attack modeling includes helping them to identify potential attacks such as unauthorized access to personal data, which indirectly helps focus on privacy \cite{ChimucoFrancisco}. Overall, the framework provides a simple, text-based interface where developers need to answer a series of questions about their application. Based on the responses provided, the framework generates detailed reports outlining security requirements, best practices (e.g., guidelines for secure coding, authentication, encryption, etc.), attack models (e.g., details about potential attacks), and test specifications (e.g., security testing approaches such as SAST, DAST, etc.). 

Separately, Giffin et al. \cite{GiffinDaniel12} proposed a web framework called Hails, designed to enhance security and privacy in web platforms that rely on third-party applications. In other words, it is designed for developers who are building web platforms like Facebook that allow third-party applications to extend their functionality \cite{GiffinDaniel12}. One of the main challenges with third-party applications is their unrestricted access to user data: once access to the user data has been granted by the platform, there are no enforced restrictions on how they handle that data, leading to potential misuse or leaks \cite{GiffinDaniel12}. For example, on a platform like Facebook, a third-party app might gain access to a user’s friend list and then share that data with advertisers without the user’s knowledge. Therefore, these platforms need to ensure that third-party applications cannot misuse or improperly access user data \cite{GiffinDaniel12}. Hails mitigates this risk by introducing a new architecture called MPVC (model-policy-view-controller), which is an extended version of MVC by incorporating security policies at the framework level \cite{GiffinDaniel12}. Hails allows platform developers to define security and privacy policies in a way that is separate from the application logic \cite{GiffinDaniel12}. For instance, in GitStar, a code-hosting platform built with Hails \cite{GiffinDaniel12}, a third-party Code Viewer app can access a user’s private repository. However, Hails makes sure that the app only shows the code to people who are specifically allowed to see it, such as collaborators. In that way, unlike conventional systems where security and privacy policies are deeply connected with application logic, Hails allows platform developers to define and enforce privacy policies separately within the platform, simplifying data access control and policy management \cite{GiffinDaniel12}. 

\subsubsection{Secure \& privacy-preserving coding}
\label{subsubsec:rivacy-preserving-coding}

In privacy-preserving application development, detecting code defects is crucial since even minor coding defects may lead to unauthorized access, data leaks, and privacy violations. For example, SQL injection occurs when attackers use vulnerable user input to manipulate database queries, allowing unauthorized access to sensitive data. Android Studio comes with some tools, such as Android Lint, to support developers in preventing program applications with privacy vulnerabilities \cite{nguyen17}. However, development issues such as requesting more permission than needed \cite{felt11, adrienne11}, incorrect use of cryptographic APIs \cite{egele13}, store sensitive information in non-private locations \cite{william11}, prove that the available developer support is not sufficient to make secure and privacy-preserving applications \cite{nguyen17}. Developers' lack of knowledge and time constraints may be a reason for that, since they also lead to privacy vulnerabilities in applications \cite{yasemin16, nguyen17, felt11, adrienne11, egele13, william11, miryung04}. For instance, due to the limited time available to develop the applications, developers tend to copy and paste some code to solve security and privacy-related problems \cite{miryung04}, which may raise security and privacy issues. A common example is reusing code snippets from Stack Overflow \cite{stackoverflow} to implement password hashing that does not have a secure hashing algorithm like Bcrypt. 

Considering the above challenges, to support developers in writing secure and privacy-preserving code, Nguyen et al. \cite{nguyen17} proposed an Android plugin called FixDroid by improving Lint’s limitations. The Android Lint tool has a feature to highlight insecure code, but with a drawback that it uses the same highlighting for all types of warnings \cite{nguyen17}. As a solution, to get developers' attention, FixDroid implemented its user interface based on insights from usable security and privacy research as well as the highlighting feature \cite{nguyen17}. Additionally, Lint provides only textual tooltips, whereas FixDroid offers code snippets with quick fixes and detailed explanations, enabling developers to easily turn insecure code into secure
code \cite{nguyen17}. Further, Lint’s lightweight code analysis technique cannot detect complex code issues \cite{nguyen17}. Therefore, FixDroid has been implemented with more complicated code analysis techniques \cite{nguyen17}. Further, if a developer copies a code snippet from somewhere, FixDroid detects its vulnerabilities using an online database of insecure code snippets \cite{nguyen17}. 

Separately, to detect internal code-level privacy vulnerabilities such as unauthorized logging (i.e., recording) of sensitive user data of large-scale software systems (e.g., WhatsApp), an automatic tool called PrivacyCAT has been introduced \cite{privacycat}. PrivacyCAT is used to detect privacy vulnerabilities before they reach the production level \cite{privacycat}. This is because accessing production traffic and fixing privacy issues may raise further privacy issues because it contains actual user information \cite{privacycat}. PrivacyCAT uses both dynamic and static taint analysis to track the flow of data and detect any data leaks \cite{privacycat}. The static analysis is performed on the source code, while dynamic analysis is performed in isolated, sandboxed environments with artificially generated sensitive data. It generates synthesized but realistic data to track how sensitive data flows through the system, ensuring that vulnerabilities are detected during development rather than after deployment \cite{privacycat}. By tracking the flow, it monitors how the system processes data at data sinks (e.g., API end-points) and exchanges through APIs to identify potential leakages \cite{privacycat}. Two years of results of industrial deployment of PrivacyCAT in WhatsApp reports that it effectively finds privacy vulnerabilities and helps prevent them at early stages of development \cite{privacycat}. 

\subsection{Manage \& Secure Personal Data}
\label{subsec:privacy-personaldata}

According to privacy laws, personal data can be referred to as Personally Identifiable Information (PII), which can be used to identify a specific person \cite{privacystreams}. Developers access and use personal data to create functional, rich, and meaningful applications \cite{privacystreams}. Mobile platforms like Android and iOS implement permission-based access control to regulate personal data usage, requiring applications to obtain user consent before accessing data \cite{privacystreams}. However, from the users' perspective, the granularity and the purpose of data access are important for users to trust the application behavior \cite{privacystreams}. This poses a challenge: "Will the developer be able to provide such granularity?". Asking developers to provide such fine-grained information through a privacy policy requires extra effort, and developers might not have the benefits and incentives (e.g., money) to put that effort into it \cite{privacystreams}. Further, the availability of different APIs and data structures makes it difficult for developers to access and use personal data \cite{privacystreams}, as they may have to follow different protocols (e.g., HTTP, WebSockets), security requirements (e.g., Jason Web Tokens, API keys), data structures (e.g., JSON, XML), etc., to manage the data effectively. 

As a solution, an Android library that comes as a functional programming model, PrivacyStreams, has been introduced to access and process different types of personal data available on smartphones, providing such granularity to users \cite{privacystreams}. Given that PrivacyStreams is provided as a library, developers can import it and make use of its APIs in the code to request personal data, process it, and take actions on it by centralizing data-handling steps at a single location \cite{privacystreams}. PrivacyStreams includes a static code analyzer that extracts application data processing steps to automatically generate privacy descriptions, as well as to analyze the granularity of data processing \cite{privacystreams}. For example, PrivacyStreams helps to expose the granularity of data processing by showing (i.e., generating privacy descriptions) whether an app is using fine-grained data (e.g., exact GPS coordinates) or coarse-grained data (e.g., city-level location). The generated description may be as follows: "This app requests LOCATION permission to get the city-level location.".

Most of the applications that provide location-based services represent the location using latitude and longitude \cite{MurukannaiahPradeep}. However, users can be provided an improved experience if users' activities and social context can be incorporated with physical location \cite{MurukannaiahPradeep}. For example, consider a coffee shop as the place and consider factors like the type of coffee shop, the atmosphere, the people that the user interacts with, and the activities the user is doing. However, implementing such personalized experiences for users without giving them unnecessary burdens is a challenge, as it is a challenge for developers to determine each user's preferences \cite{MurukannaiahPradeep}. 

This motivates Murukannaiah and Singh to propose a framework called Platys to develop place-aware applications \cite{MurukannaiahPradeep}. Here, "place" refers to the combination of factors discussed above, other than the location \cite{MurukannaiahPradeep}. Platys extends location-based applications beyond location coordinates by incorporating user-specific contextual information \cite{MurukannaiahPradeep}. The core of Platys is its middleware, which consists of a machine learning technique called active learning to learn a user-specific model for places \cite{MurukannaiahPradeep}. The middleware exposes the learned model to applications at runtime, allowing developers to use it to develop place-aware applications without making efforts to connect with low-level sensors (e.g., GPS sensors, accelerometer, gyroscope, etc.) \cite{MurukannaiahPradeep}. Also, since Platys runs locally on users' devices, it ensures the user's privacy \cite{MurukannaiahPradeep}. A developer study proved that the Plays framework reduces the developer's effort to build place-aware applications \cite{MurukannaiahPradeep}.

\subsection{Address Vulnerabilities in Third-party Dependencies}
\label{subsec:privacy-library}

Integration of third-party libraries and SDKs, in other words, dependencies, makes the development process easier by providing reusable functionalities and accelerating the development process \cite{nguyen20, ekambaranathan23, enisa2018, balebako14}. However, integrating these dependencies introduces privacy and security risks, particularly if the libraries are outdated or improperly configured \cite{nguyen20}. As a result, previous studies have examined the security and privacy risks associated with third-party libraries and provided them with solutions, which we discuss in this section \cite{ekambaranathan23, nguyen20}. 

When an application relies on dependencies, they should be updated on time regularly to overcome the problems of outdated dependencies \cite{nguyen20}. For example, Heartbleed is a vulnerability of an outdated version of the OpenSSL cryptography library, allowing attackers to steal sensitive data from servers \cite{heartbleed}. Updating it may ensure protection against known threats. However, developers often fail to update them due to concerns such as extra effort, fear of incompatibilities, and lack of awareness of updates \cite{backes16, nguyen20, derr17}. To address this, Nguyen et al. \cite{nguyen20} proposed a tool called Up2Dep that analyzes the libraries used in the code base and helps developers avoid insecure library versions by providing them with feedback about outdated libraries \cite{nguyen20}. Further, it warns developers about vulnerabilities and cryptographic API misuse of libraries, allowing them to make informed decisions before integrating a library \cite{nguyen20}.

Beyond being up-to-date with third-party dependencies, privacy implications in third-party libraries and SDKs should also be considered, specifically when developing children's applications \cite{ekambaranathan23}. From the developers' perspective, they face difficulties such as identifying relevant privacy-compliant third-party libraries and SDKs from the extensive range \cite{konrad21, ekambaranathan23, ekambaranathanzhao21}, configuring and managing SDK settings correctly \cite{ekambaranathanzhaovan, ekambaranathan23}, and finding privacy-friendly alternative libraries and SDKs \cite{ekambaranathanzhao21, ekambaranathan23}. Further, understanding the data-handling process of third-party libraries and SDKs is also difficult due to their lower transparency about the data-handling processes \cite{binnszhao18, ekambaranathan23, myrstad2021}. 

Considering these challenges, Ekambaranathan et al. \cite{ekambaranathan23} proposed a web-based tool called DataAvalanche.io to support developers navigating legal issues and privacy implications associated with third-party libraries and SDKs. Developers can use this tool to get more privacy information about commonly used libraries and SDKs, including data trackers associated with them \cite{ekambaranathan23}. Additionally, the proposed tool provides clearer instructions to configure child-specific SDKs and provides a list of privacy-preserving alternative SDKs with documentation of their data-handling practices  \cite{ekambaranathan23}. A key feature of the tool is its ability to visualize with which countries the SDKs share data, allowing developers to assess whether an SDK complies with regional privacy regulations like GDPR \cite{ekambaranathan23}. 

\subsection{Make Software Privacy Compliant and Transparent Using Privacy Statements}
\label{subsec:privacy-statements}

This section discusses the developer support available in the existing literature to create privacy statements. We discuss this under three different groups based on the support they provide for developers: privacy label generation \cite{gardner22, tianshi24}, privacy policy generation \cite{zimmeck2021PrivacyFlashPA, coconut}, and privacy notice generation \cite{tianshi21}. 

\subsubsection{Privacy label generation}
\label{subsubsec:privacy-label-generation}
Privacy labels are short notices about how applications handle user data, which are often displayed when an application is listed on the app store \cite{kelley09}. However, when implementing them in software development practices, developers face some challenges due to a lack of knowledge about privacy concepts (e.g., data minimization), misunderstanding about technical terms used in privacy labels (e.g., precise, coarse, etc.), and difficulties in privacy label generation when using third-party libraries with black-box data handling practices \cite{gardner22, tianshi24, jain23}. Further, differences in privacy labels and the actual application behaviors will lead to losing users' trust in applications \cite{jain23} and fines for developers for privacy violations \cite{gdpr16}.  

To assist developers in creating accurate privacy labels for Google Play, an Android Studio plugin called Matcha was introduced \cite{tianshi24}, while Privacy Label Wiz (PLW) was developed for Apple iOS applications \cite{gardner22}. Both use static code analysis techniques to analyze the data-handling practices of the application \cite{tianshi24, gardner22}. Matcha automatically detects third-party SDKs used in the codebase and helps fill parts of privacy labels based on the information provided by the SDKs' privacy documentation \cite{tianshi24}. 

On the other hand, PLW utilizes static code analysis techniques to scan the application codebase based on iOS permissions used in the Swift code \cite{gardner22}. Since some permissions don't align with Apple's data types, it informs developers about potential misalignments regarding data types \cite{gardner22}. For example, let's say an iOS app uses the "Camera" permission to allow users to take photos. In this case, the app might not explicitly collect or store photos. Therefore, the "Camera" permission could be mapped to Apple's "Photos or Videos" data type.  Additionally, PLW prompts developers with questions asking developers to confirm or refine the labels, ensuring the final set of privacy labels is an accurate reflection of the application behavior \cite{gardner22}. 

Matcha has another facility to automatically generate an XML file that consists of all data collection and sharing, and gives the ability for developers to review it and add manual annotations if required \cite{tianshi24}. Then, Matcha will generate a CSV file consisting of labels that later can be uploaded into the Google Play developer console \cite{tianshi24}. 

Apart from those plugins (i.e., tools), Jain et al. \cite{jain23} proposed a novel methodology called "fine-grained localization of privacy behaviors" to accurately identify privacy behaviors (i.e., how an application uses personal information and why) in source code and predict privacy labels. It helps developers create high-quality privacy labels by reducing the time and mental effort required by automating the privacy-label generation process \cite{jain23}. At the same time, the study demonstrates high accuracy in predicting privacy labels \cite{jain23}. Since this approach uses individual statements in the source code for the analysis process by going beyond the class or method level, it avoids the noise and ambiguity that come with analyzing larger code blocks \cite{jain23}. It eventually improves the accuracy of the generated labels \cite{jain23}.

\subsubsection{Privacy policy generation}
\label{subsubsec:privacy-policy-generation}
The privacy policy is a way to convey the application's privacy practices with details such as how the user's and device's data is collected, used, shared, and deleted \cite{google_play_agreement, apple_developer_agreement}. However, developers face a challenge in accurately representing how data is collected, used, shared, and deleted \cite{andow20, zimmeck2021PrivacyFlashPA} because of having an inaccurate understanding of the application behavior \cite{coconut}. It happens especially when third-party APIs or libraries used in the application have a lack of documentation about their data-handling practices \cite{coconut}. This issue causes developers to produce inaccurate privacy policy documents. Further, in some situations, multiple data sources are controlled by one permission \cite{privacystreams}. For instance, in a situation where developers need the user's name, once the user grants permission, it lets developers access even emails other than the name. These barriers result in producing privacy policies that may not convey the actual data-handling practices to users and may not fully protect users' data or comply with regulations. 

In our SLR, we have identified two developer-supporting tools, PrivacyFlash Pro \cite{zimmeck2021PrivacyFlashPA} and Coconut \cite{coconut}, that have been proposed to support developers generating privacy policies. 

PrivacyFlash Pro \cite{zimmeck2021PrivacyFlashPA} has been implemented for iOS applications coded in Swift and integrated Swift and also for libraries coded in Objective-C \cite{zimmeck2021PrivacyFlashPA}. PrivacyFlash Pro combines static code analysis techniques and questionnaire-based wizards to generate privacy policies \cite{zimmeck2021PrivacyFlashPA}. Fully relying on questionnaire-based policy generators may result in inaccurate policies because the generation is fully dependent on the developers' answers, which may not align with actual data practices of the application \cite{zimmeck2021PrivacyFlashPA}. PrivacyFlash Pro is a macOS desktop application that consists of a code analysis logic written in Python and a user interface developed in JavaScript \cite{zimmeck2021PrivacyFlashPA}. 

On the other hand, Coconut uses an annotation-based technique, and developers need to annotate personal data practices used within the application while coding in a predefined format \cite{coconut}. This process encourages developers to explicitly document how and why personal data is collected, used, and shared \cite{coconut}. Coconut provides real-time feedback on potential privacy issues, such as violations of privacy principles or unnecessary data collection \cite{coconut}. Additionally, it provides quick fixes to ease the development process, such as generating annotation skeletons with automatically filled fields (e.g., automatically filling the data type) and recommending more privacy-friendly alternatives (e.g., using coarse-grained location data instead of fine-grained when high precision is not required) \cite{coconut}. These features help developers adhere to recommended privacy practices while reducing the cognitive workload of manually identifying and addressing privacy concerns \cite{coconut}. Coconut combines all the annotations into a summary panel, which provides a comprehensive overview of the app's personal data practices \cite{coconut}. This summary can be directly translated into a privacy policy. 

\subsubsection{Privacy notice generation}
\label{subsubsec:privacy-notice-generation}
In-app privacy notices provide real-time, brief, and contextual messages about users' data handling when they interact with the applications. However, developers face some challenges when they create privacy notices. One of the main reasons is the lack of awareness among developers about the privacy concept \cite{tianshi21} and the design choices and best practices when crafting privacy notices \cite{nouwens20}, which may lead to poor communication and failure to inform users about how their data is handled in the application. When application development relies on third-party libraries, the design and implementation of in-app privacy notices may further complicate developers because developers are not always aware of third-party applications' data-handling practices when they are not documented \cite{balebako14, coconut}. Additionally, developers prioritize security over privacy \cite{hadar18}, focusing more on preventing data breaches rather than ensuring users' data-handling transparency.

We found a paper that highlights these challenges and proposes an Android Studio plugin called Honeysuckle \cite{tianshi21}, which is a programming tool proposed by Tianshi Li et al. to help developers generate in-app privacy notices. Honeysuckle utilizes an annotation-based approach to generate privacy notices where developers need to annotate each data source (where data is collected) and data sink (where data is sent or stored), describing associated data practices \cite{tianshi21}. For example, if an app collects the user's location, the developer may annotate the code where the location data is collected and specify the purpose of the data collection (e.g., "for providing nearby restaurant recommendations"). Also, developers can use a configuration file at programming time to fine-tune privacy notices \cite{tianshi21}. This configuration file allows developers to customize aspects such as the context information that should be displayed in the notices and the behavior of the notices (e.g., how frequently they are shown or under what conditions) \cite{tianshi21}. For instance, a developer may need to configure the system to show a privacy notice only when the app accesses the user's location in the background. Honeysuckle helps developers to generate privacy notices that are more contextualized and user-centered without needing to manually implement each notice from scratch \cite{tianshi21}.

\subsection{Improve Privacy Education and Awareness}
\label{subsec:privacy-education}

Privacy-by-design (PbD) emphasizes integrating privacy into software from the beginning rather than taking it as an afterthought \cite{cavoukian2012operationalizing}. However, developers face challenges because of its broad interpretation \cite{ShapiroStuart10, TangYing21}. For example, PbD suggests minimizing data collection broadly, but it does not specify how to achieve it (i.e., there is no proper technical implementation) in software development. Similarly, regulations like GDPR and CCPA lack clear guidance on technical implementation \cite{TangYing21, EarpJulia, CompagnaLuca, ShapiroStuart10, weirCharles23}. For instance, GDPR's storage limitation principle requires that personal data should not be kept longer than necessary. However, it does not specify exact retention periods or how developers should enforce deletion. Because of this interpretation and implementation issues of PbD and regulations, enhancing privacy education among developers may be beneficial.

Privacy Ideation Cards (PIC), introduced by Luger et al. \cite{LugerEwa15}, help software developers understand how PbD principles and privacy regulations apply to software development. Tang et al. \cite{TangYing21} utilized PICs and evaluated practically how PICs can be useful for software engineering students as an educational tool to help them learn privacy. PIC contains a deck of cards with details about PbD and privacy regulation principles (e.g., data minimization, explicit consent, etc., with their detailed explanation). In an ideation session, students draw PIC cards and engage in discussions focusing on how the PbD and regulatory principles on the cards apply to their real-world software projects \cite{TangYing21}. The evaluation results revealed that students engaged in deeper discussions about specific aspects of the project from the privacy perspective \cite{TangYing21}. 

Another approach, a workshop-based intervention proposed by Weir et al., addresses the challenge that developers face when effectively communicating security and privacy concerns to product managers who have limited awareness of security and privacy risks \cite{weirCharles23}. The product manager is responsible for making decisions related to time and resource allocation to address security and privacy concerns \cite{weirCharles23}. However, the responsibility of implementing security and privacy into software is on the development team \cite{AlhirabiNada23}, implying that it is important to help them communicate when making security and privacy decisions \cite{weirCharles23}. This intervention includes structured workshops to improve the developers' awareness about identifying and prioritizing security and privacy issues and then framing these issues in terms of business benefits to communicate with product managers \cite{weirCharles23}. The evaluation results revealed that this workshop motivated developers to engage with product managers in security and privacy-related decision-making \cite{weirCharles23}.

\section{Discussion} \label{discussion}

In the discussion section, we discuss the limitations of the identified solutions (RQ2) that we discussed in the results section (i.e., Section \ref{results}) and suggest potential future improvements and future research avenues under the same themes as we reported the results. 

\subsection{Consider Privacy in the Early-stage of Development}
\label{subsec:dis-privacy-requiremnt}

In Section \ref{subsec:privacy-requirement}, we discussed different approaches that support developers integrating privacy in the early stage of software development. The Privacy Criteria Method (PCM-tool) \cite{PeixotoMarianaPCMTool, PeixotoMarianaEvaluating, PeixotoMariananatural}, Restricted Misuse Case Modeling (RMCM) \cite{MAI2018165}, and collaborative threat modeling and identification techniques such as LINDUNN-Go \cite{WuytsKim20} and ThreatPoker \cite{RyggeHanne} offer structured methods to embed privacy into software requirements, design, and development. However, how far these approaches have addressed the challenges that developers face in early-stage privacy integration should be discussed.

\subsubsection{Requirement specification.} PCM-tool was developed to be used in an agile environment to provide developers with structured guidance for privacy requirement specification  \cite{PeixotoMarianaPCMTool, PeixotoMarianaEvaluating, PeixotoMariananatural}. However, evaluation carried out with industry practitioners revealed that the suitability of PCM, especially in agile, is questionable \cite{PeixotoMariananatural}. One potential improvement is integrating the PCM tool with agile project management tools like Jira \cite{atlassian_jira}, where the requirements are specified throughout the project cycle. Integrating PCM with Jira-like software may increase the usability of PCM (i.e., increase the user experience (UX)) since managing a separate tool only for privacy requirements may be an extra overhead for the developers. A key limitation of the PCM tool is that it requires the developer's manual intervention to specify the legal basis of privacy requirements (e.g., GDPR) \cite{PeixotoMarianaEvaluating, PeixotoMariananatural, PeixotoMarianaPCMTool}. However, since these specifications depend on the knowledge of the developer who does this activity, it may lead to errors (e.g., developers with limited privacy knowledge may provide an incorrect legal basis). Therefore, implementing an intelligent system to cross-check the correctness of the specified legal basis and warn the team if any misalignment is encountered may significantly enhance the accuracy of the privacy requirements' regulatory compliance. 

Similarly, we discussed Restricted Misuse Case Modeling (RMCM), which is a structured method for defining privacy and security threats through use case and misuse case diagrams \cite{MAI2018165}. However, it is required to have privacy expertise to use it \cite{MAI2018165}, hindering its practical applicability among developers who have limited privacy expertise. Since not all organizations have that expertise, integrating RMCM with development environments (e.g., IDE) and then guiding developers through automated tools to identify misuse cases may be a potential improvement. 

\subsubsection{Threat modeling \& identification.} Beyond privacy requirement specification, threat modeling and identification is a crucial technique for privacy risk assessment, which is done in the early stage of software development \cite{WuytsKim20}. LINDUNN-Go was developed as a simplified collaborative card-based toolkit for threat identification \cite{WuytsKim20}. Both LINDUNN and LINDUNN-Go cover 7 threat categories. However, the LINDDUN-GO toolkit focuses on the most common or critical threats within each category, while LINDUNN focuses on every possible threat scenario. This implies that the effective applicability of the toolkit in industry practice is still questionable. For example, some complex applications (e.g., healthcare applications) may contain specific privacy threats that are not covered by the most common or critical threats. Therefore, more evaluations are required to validate the effectiveness in industrial applications \cite{WuytsKim20} since they may be complex and need more threat scenarios than LINDUNN-Go covers. 

Even though ThreatPoker did not explicitly discuss its limitations, we propose some future experiments based on the findings of its evaluation. The evaluation was conducted with students. However, because the knowledge levels (i.e., privacy and security knowledge) of industry practitioners and students can differ, applicability in the industry cannot be guaranteed based on the current results. Subjectivity may also apply when identifying threats and takes time to make a final decision as a team. Because industry projects may have specific timelines, ThreatPoker should be tested in the industry to determine how subjectivity affects results. As a result, ThreatPoker also requires further empirical evaluations to determine its effectiveness in industry settings \cite{RyggeHanne}.

Further, since both LINDUNN-Go and ThreatPoker are card-based games, the usability may be lower when the team consists of a large number of members. Further, playing card games may not be effective in time-constrained software development environments. Therefore, converting these games to digital versions (e.g., software applications) may enhance their usability as well as their scalability. 

\subsection{Consider Privacy During the Design and Development Stages}
\label{subsec:dis-integrate-privacy}

We discussed the available developer support in the current literature to embed privacy during the design or development processes in five different categories, as shown in Figure \ref{results}. Below, we discuss the implications of the results and potential future directions based on their limitations.

\subsubsection{Tool-based design and development}

Developers in small teams, small and medium-sized enterprises (SMEs), or individual developers engage in both design and development activities regardless of their position in the team \cite{RivasLornel08}. For example, software developers may be involved in IoT design activities. The identified solutions, Parrot \cite{AlhirabiNada23} and Canella \cite{AljeraisyAtheer24}, targeted such teams. Parrot specifically highlighted that it can be used by software developers, senior software developers, or even architects who are involved in the IoT application design process \cite{AlhirabiNada23}.

However, given that Parrot's primary regulatory framework is GDPR \cite{AlhirabiNada23}, its applicability in applications that operate across multiple jurisdictions is questionable. For example, developers may want to build an IoT application that processes users' information who are in Europe and Australia. It necessitates adhering to both GDPR \cite{gdpr16} and the Australian Privacy Act \cite{apa88}. Therefore, future implementations should incorporate other regulatory compliances such as CCPA \cite{ccpa18}, Australian Privacy Act \cite{apa88}, and the New Zealand Privacy Act \cite{nzpa20} to ensure that developers can adhere to regulatory requirements based on their target markets. Additionally, it was developed for healthcare applications where privacy is critical \cite{AlhirabiNada23}. It may be challenging to transfer it to other domains \cite{AlhirabiNada23}. Therefore, extended evaluations are needed to evaluate the usability in other domains.

Further, some usability concerns of Parrot have been identified in the evaluation process, highlighting that developers need more responsive feedback mechanisms \cite{AlhirabiNada23}. For example, if a developer selects a certain data type to be used in the application (e.g., location data), the tool might not immediately show how this selection affects privacy compliance or data security (e.g., the feedback might be delayed or not detailed enough). Improving the user interface (UI) and user experience (UX) by offering quick context-aware suggestions (i.e., tailored to the specific context of the IoT application) and providing pre-configured templates for common IoT use cases may significantly enhance its effectiveness. 

Similarly, Canella's limitations impact its usability \cite{AljeraisyAtheer24}. Developers raised concerns about its unintuitive interface, particularly due to the small size and colorful blocks \cite{AljeraisyAtheer24}. For instance, some developers missed Canella's warning due to their small size and colorful blocks \cite{AljeraisyAtheer24}. While visual programming environments may be beneficial for beginner-level developers, experienced developers may prefer traditional code-based approaches. Therefore, incorporating both block-based and text-based development into Canella will allow developers to choose the best option and may increase Canella's usability. 

Further, similar to Parrot, Canella's limited regulatory compliance support may affect when applying it in the broader IoT development domains \cite{AljeraisyAtheer24}. Therefore, increasing the support for more privacy regulatory frameworks may make Canella more robust for real-world deployment.

Finally, since both Parrot and Canella educate developers about best privacy-ensured design and development practices \cite{AlhirabiNada23, AljeraisyAtheer24}, expanding them as educational tools by adding interactive tutorials and exercises may educate developers about practical privacy implementations, as we discussed in Section \ref{subsec:dis-privacy-education}. It will help developers to stay in one tool without relying on separate educational tools.

\subsubsection{Guideline-based design and development}
\label{guideline-based-design-development}

In Section \ref{subsubsec:privacy-integrate-guideline}, we discussed sets of guidelines that are privacy-oriented software development (POSD)\cite{BaldassarreMaria} and PbD-based IoT designing guidelines \cite{PERERA2020238} to help developers integrate privacy into software development. However, their practical effectiveness is questionable because of reasons such as requiring manual intervention, taking too much time to understand, offering limited programming-language support, and being evaluated only for re-engineering a system \cite{BaldassarreMaria, PERERA2020238}. 

Since POSD's support has only been evaluated in re-engineering contexts \cite{BaldassarreMaria}, it suggests that POSD primarily functions as a corrective approach rather than a proactive approach. Developing a system considering privacy as an afterthought may not be effective in both time and cost terms. Therefore, proposing POSD in a way that it is suitable for the forward mode will make this set of guidelines more usable \cite{BaldassarreMaria}. 

Fortify SCA is the proposed code analyzer in POSD \cite{BaldassarreMaria}. Even though it can analyze codes of more than twenty languages, it is in vain since the PKB of POSD supports only exporting Java-based structures for privacy patterns \cite{BaldassarreMaria}, reducing its practical usability among developers who work in languages such as Python, JavaScript, or C\#. Therefore, it can be recommended to expand PKB to support more languages to get the benefit of the code analyzer \cite{BaldassarreMaria} since the tool may remain inaccessible to many development teams without broader language support. 

Additionally, in POSD, developers need to manually identify system vulnerabilities using code analysis and then input the detected vulnerabilities to PKB \cite{BaldassarreMaria}. That process may introduce human errors. Therefore, replacing the manual workload with an automated process may reduce potential human errors. For example, machine learning-based vulnerability detection. However, they might not be 100\% accurate. Therefore, implementing a feedback loop to get the developer's feedback for the predictions and then retraining machine learning models based on the feedback may further enhance the prediction accuracy. Further, integrating it with integrated development environments (IDEs) may help developers follow the POSD guidelines while coding without making extra effort. Another issue is POSD’s incompatibility with Agile methodologies \cite{BaldassarreMaria}. Since Agile methodologies prioritize continuous integration, static privacy guidelines like POSD may not be suitable. Without evaluations in real-world Agile environments, the suitability of POSD in Agile environments is unclear. Therefore, the applicability of guidelines in an agile environment should also be evaluated. 

On the other hand, the usability of the proposed PbD-based IoT guidelines is less because the way the authors conveyed the guidelines (i.e., printed sheets) for developers was difficult to follow and time-consuming \cite{PERERA2020238}. As a solution, Privacy Ideation Cards (PICs) may be effective for developers to quickly familiarize themselves with the guidelines since PICs have shown their effectiveness in practice \cite{PERERA2020238, TangYing21, LugerEwa15}, as we discussed in section \ref{subsec:privacy-education}. Similarly, providing developers with an abstract version of guidelines, which are privacy tactics and patterns, may improve their usability \cite{PERERA2020238}, especially among developers who have less privacy knowledge and who face difficulties in following guidelines. Further, even though utilizing physical cards (e.g., PIC \cite{TangYing21, LugerEwa15}) has shown effectiveness, it may be more effective if a user-friendly interface can be developed using human-computer interaction techniques for developers to follow during their design process \cite{PERERA2020238}. 

\subsubsection{Methodology-based design and development}

The identified methodology-based approaches, Privacy by Evidence (PbE) \cite{PedroBarbosa}, CryptSDLC \cite{LoruenserThomas18}, and Model-Driven Development (MDD) for GDPR compliance \cite{krstic24}, were discussed in detail in Section \ref{subsubsec:privacy-integrate-method}. However, their effectiveness in practice should be further discussed because of their limitations, such as complexity, reliance on manual processes, and limited real-world validation \cite{PedroBarbosa, LoruenserThomas18, krstic24}. 

As we discussed, PbE relies on extensive documentation, but it is time-consuming \cite{PedroBarbosa} and may be impractical for rapid development processes where speed is prioritized. For instance, in agile or DevOps workflows, developers often need to deliver features quickly, and the manual effort required to write detailed privacy-related artifacts (e.g., risk assessments, compliance proofs, and attack simulation reports) may slow down the development process. This makes PbE potentially impractical for teams that are operating under tight deadlines or with limited resources. 

Even though PbE was proposed to address the lack of clear implementation guidance for PbD \cite{PedroBarbosa}, it may not provide a simplified privacy integration for developers with limited privacy expertise. For example, developers may be unfamiliar with privacy models like k-anonymity or differential privacy. As a result, they may struggle to apply these concepts correctly without having proper training or support. Therefore, after a careful analysis of PbE, we suggest developing automated or semi-automated tools to assist developers during the PbE process to reduce the manual intervention of developers and streamline the PbE process. For example, developing a tool that can scan the codebase for sensitive data handling and flag areas where privacy techniques such as anonymization or encryption are needed. It may help to increase awareness about privacy concepts among developers. Further, integrating them with IDEs may streamline the development process.

CryptSDLC \cite{LoruenserThomas18} was proposed as a solution to the technical complexity of cryptographic engineering. It does not explicitly discuss privacy. However, it lays a strong foundation for privacy-preserving software development. For instance, CryptSDLC emphasizes tools and primitives to ensure the confidentiality of the system \cite{LoruenserThomas18}, which directly supports privacy by protecting personal data from unauthorized access and disclosure. However, the feasibility of CryptSDLC is demonstrated only in the field of cloud computing \cite{LoruenserThomas18}. Therefore, we suggest the standardization of cryptographic primitives and their properties in CryptSDLC to enhance communication and interoperability between different systems and stakeholders, making it easier to adopt CryptSDLC in diverse software fields. For example, standardized cryptographic protocols may enable seamless integration of privacy-preserving features across different areas such as IoT devices, healthcare, or financial systems, ensuring consistent data protection. Further, standardizing cryptographic primitives may create a common vocabulary (i.e., common standardized cryptographic primitives such as AES-256 encryption or SHA-3 hashing, irrespective of the domain) that all stakeholders who work on different projects can understand. Additionally, integrating automated tools to support the CryptSDLC process \cite{LoruenserThomas18} may streamline the design process and reduce human errors. It may help to reduce the likelihood of human errors and speed up the development process. For example, automatic tools that generate secure configurations for cryptographic libraries and validate that the libraries meet the required security and privacy requirements.

Further, the organization's privacy culture is improved with proper communication among the team members as per the survey conducted by Tahel et al. \cite{TahaeiMohammad21}, which eventually positively affects project development. The architecture used in CryptSLDC facilitates communication and collaboration among stakeholders such as cryptographers, software developers, and application designers \cite{LoruenserThomas18}, which may help to build the organization's privacy culture. 

Next, we discussed a Model-Driven Development (MDD) methodology \cite{krstic24}, which aimed to address the challenge that developers face when developing systems specifically with purpose limitation and user consent, as mandated by GDPR. However, it is a simplified interpretation of GDPR. Extending its compliance scope beyond purpose limitation and user consent may make the methodology more impactful and scalable. Further, since the model transformation is supported only for C\# and Python web applications \cite{krstic24}, it covers a small portion of the development stacks. This may limit usability in large-scale systems where developers work with multiple programming languages. Therefore, extending this to other programming languages such as JavaScript for web applications, Java, and Dart for mobile applications may broaden the applicability of the methodology. 

\subsubsection{Framework-based design and development}

Secure SDLC Framework \cite{KangSooyoung}, secD4CloudMobile \cite{ChimucoFrancisco}, and Hails \cite{GiffinDaniel12} are frameworks we discussed in Section \ref{subsubsec:privacy-integrate-framework} that attempted to enhance security and privacy integration in software development.

First, the Secure SDLC Framework \cite{KangSooyoung} primarily focuses on security rather than privacy. Even though its primary focus is security, which is centered around the fundamental security principles of confidentiality, integrity, and availability \cite{KangSooyoung}, privacy is a subset of these. However, privacy-by-design (PbD) principles are not explicitly incorporated \cite{KangSooyoung}, limiting their effectiveness during the early stage of software development. Additionally, it consists of some manual mappings of secure SDLC standards and guidelines \cite{KangSooyoung}, making it more resource-intensive and possibly not possible to apply in large-scale projects with limited time and money constraints. Therefore, automating that process and integrating PbD principles may reduce the developer burden and improve its efficiency. Further, we recommend implementing a feedback loop to collect information from stakeholders such as developers, security, and privacy teams in order to stay up-to-date on emerging threats and understand how competitors have addressed them. Since this may improve the communication among stakeholders, it may improve the organization's security and privacy culture, as discussed in section \ref{subsec:dis-privacy-education}. 

In a related effort that does not specifically provide privacy support, Chimuco et al. proposed another framework targeting developers who work on cloud and mobile ecosystems \cite{ChimucoFrancisco}. Its validation was conducted on a small scale \cite{ChimucoFrancisco}, limiting confidence in its effectiveness for real-world application. Therefore, further evaluations are needed to check the effectiveness in industry settings. Additionally, incorporating AI with all the modules of the framework has been emphasized by the authors \cite{ChimucoFrancisco}. For instance, to improve the accessibility of the text-based console in the framework, a natural language processing-based developer interaction has been proposed \cite{ChimucoFrancisco}. By integrating NLP, developers may be able to interact with the framework using natural language (e.g., typing or speaking in plain English) instead of navigating through a structured questionnaire. Further, the current framework relies on a rule-based system that generates recommendations based on predefined inputs \cite{ChimucoFrancisco}. However, NLP can enable the framework to understand the context of the developer’s queries and ensure that the framework provides more tailored recommendations and relevant guidance.

Next, we discussed Hails, where the authors highlighted the risk of data privacy breaches in current web platforms like Facebook that do not have restrictions on how third-party applications handle user data \cite{GiffinDaniel12}. Hails is built using the Haskell language, which is not as widely used as other languages like JavaScript or Python \cite{GiffinDaniel12}. It may pose a barrier to adoption for some platform developers. For instance, a platform developer familiar with Python or JavaScript might find it challenging to transition to Haskell. Haskell’s ecosystem and libraries for web development are less mature and less widely documented compared to those in more mainstream languages. That makes it harder for newcomers to find resources and community support. Additionally, the authors highlighted that the query system in Hails is simple, which hinders performance in situations where complex queries are needed \cite{GiffinDaniel12}. For example, if a platform like GitStar \cite{GiffinDaniel12} needs to filter, aggregate, or join large datasets for a scenario like finding all repositories owned by users in a specific region, Hails’ current query system may not be sufficient. Therefore, we suggest extending the Hails framework to support more queries, such as filtering, aggregations, joins, and techniques for optimizing query execution, such as indexing and caching. Furthermore, Hails does not provide adequate support for platform developers to become familiar with the framework. For example, developers may struggle to write code from scratch for common issues related to security and privacy policies (e.g., access control mechanisms). Therefore, we suggest developing a tool that can generate boilerplate codes for common application components, as well as a debugging tool to debug security and privacy policy issues. Additionally, integrating Hails with popular IDEs (e.g., Visual Studio Code or IntelliJ) will provide better code completion, linting, and debugging support.

\subsubsection{Secure \& privacy-preserving coding}
\label{sec:subsub:seccode}

In our SLR, we found two papers that proposed developer support to write secure code, FixDroid \cite{nguyen17}, and PrivacyCAT \cite{privacycat}. The primary focus of both solutions is to support writing secure code, but FixDroid provides general code security, while PrivacyCAT specifically detects privacy defects in code \cite{nguyen17, privacycat}. 

We discussed how FixDroid was developed to enhance Android Lint by providing real-time feedback and quick fixes for developers \cite{nguyen17}. However, it was developed to be used in Android environments \cite{nguyen17}, limiting its usability for iOS, web, and other programming environments. For example, this tool may not be effective for a developer or an organization that has an application written for Android, iOS, and web platforms since the development environments and the programming languages are different. They may need to look for alternative tools (other than for Android) that are compatible with different environments and languages. Therefore, we suggest extending FixDroid for other development environments, such as iOS development, web application development, and other languages, such as C++, JavaScript, and Python, to increase usability. FixDroid does not leverage developer-specific context to provide suggestions \cite{nguyen17}. That means that FixDroid provides generic security and privacy feedback to all developers, regardless of their individual coding habits, experience level, or the specific context of the application they are building. Therefore, in the future, we suggest using machine learning to provide personalized developer support utilizing developer-specific contexts such as the developer's coding style, level of experience, historical security and privacy issues, and the context of the application (e.g., healthcare, social media). 

On the other hand, PrivacyCAT was designed to detect privacy vulnerabilities in WhatsApp's code using both static and dynamic analysis and is designed to run on both WhatsApp client and server code \cite{privacycat}. PrivacyCAT has been evaluated only in WhatsApp \cite{privacycat}, leaving its applicability in other large-scale organizations questionable. For instance, WhatsApp’s architecture may not be the same as the structures of other applications. Therefore, its generalizability to different codebases, development teams, and organizational environments should be tested. Technically, applying machine learning along with the already used static and dynamic taint analysis to identify more patterns in privacy violations may improve the system's ability to detect new and emerging threats. Further, since this has been proposed for large organizations that handle large code bases, it may be helpful to use distributed analysis techniques to scale up PrivacyCAT to handle larger code bases distributed among different teams of the organization. 

\subsection{Manage \& Secure Personal Data}
\label{subsec:dis-personal-data}

We discussed in section \ref{subsec:privacy-personaldata} how  PrivacyStreams \cite{privacystreams} and Platys \cite{MurukannaiahPradeep} offer developer support for managing and securing personal data.

PrivacyStreams comes with a static analyzer that can be used to check how personal data is accessed, processed, and also the granularity of its usage \cite{privacystreams}. However, it cannot handle complex data operations such as machine learning (ML) transformations \cite{privacystreams}, limiting its applicability in scenarios like ML-based personalized recommendations, sentiment analysis, and user behavior predictions. Expanding PrivacyStreams to incorporate ML operations may significantly enhance its usability since it enables developers to perform advanced data processing while maintaining data usage transparency. Additionally, if we can create privacy policies, descriptions, and in-app privacy notices with the help of the static analyzer as used by the tools discussed in section \ref{subsec:privacy-statements}, it will be useful for developers to provide end-users with detailed information about how their data is handled. Also, if an application relies on third-party libraries that do not use PrivacyStreams to handle personal data, the static analyzer fails to capture complete data processing details \cite{privacystreams}. Developing a middleware layer that acts as a bridge between third-party libraries and PrivacyStreams may address this issue. This middleware should be able to intercept data requests and responses from third-party libraries and then automatically translate them into PrivacyStreams-compatible formats. Intercepting data requests can be done using API hooking (e.g., Frida \cite{frida2025}) and Aspect-Oriented Programming (e.g., Spring AOP \cite{springAOP2025}), while translating can be achieved through simple pre-defined mapping functions. Furthermore, if an application has been coded using code obfuscation (i.e., hiding) techniques, it will be a barrier for PrivacyStreams to analyze the application \cite{privacystreams}. We suggest developing a metadata layer within PrivacyStreams to allow developers to annotate the obfuscated data pipelines describing their granularity and purpose. Then, the static analyzer can rely on this metadata to reconstruct the data flow and provide data transparency. 

On the other hand, in Platys, we did not discuss the features that end-users have and their contribution to Platys functions in detail because the SLR mainly focuses on developer support. Platys utilizes active learning and semi-supervised learning methods \cite{MurukannaiahPradeep}. However, the accuracy of the trained model depends on the accuracy of the place labels provided by the users \cite{MurukannaiahPradeep}. If users provide incomplete or inconsistent place labels, the model may misclassify locations, leading to privacy risks or incorrect recommendations. Therefore, the suitability of machine learning should be evaluated again with different scenarios of users (e.g., high-mobility users, low-mobility users, privacy-conscious users, and privacy-indifferent users), and different machine learning techniques (e.g., transfer learning methods, ensemble models - random forests) can also be tested in parallel. Furthermore, the authors mentioned the high battery consumption of Platys \cite{MurukannaiahPradeep}, which remains unexplored. This is important to check because developers may be hesitant to use it in applications if it consumes too much battery power, as users may reject the application. We suggest comparing battery consumption with different complex and tiny machine-learning models to check whether it has a direct effect on the battery consumption.

\subsection{Address Vulnerabilities in Third-party Dependencies}
\label{subsec:dis-third-party}

Third-party libraries and SDKs enhance development efficiency but introduce security and privacy risks if they are outdated or misconfigured \cite{nguyen20, ekambaranathan23}, as we discussed in Section \ref{subsec:privacy-library}. Even though the tools discussed, Up2Dep \cite{nguyen20} and DataAvalanche.io \cite{ekambaranathan23}, tried to address these challenges, they have their own limitations. 

Since Up2Dep relies on LibScout \cite{LibScout} and Cognicrypt \cite{CogniCrypt} to identify API changes between different versions of libraries and cryptographic misuses of libraries, respectively, Up2Dep inherits the limitations associated with them. For example, LibScout sometimes may not work if the functionality of APIs in libraries changes while the structure stays the same \cite{nguyen20}. Integrating machine learning techniques may solve this by helping to recognize semantic changes in APIs. For example, natural language processing (NLP) models may help to analyze API documentation or code comments to detect functional changes that are not reflected in structural changes. Similarly, Cognicrypt sometimes reports false positives about cryptographic misuse \cite{nguyen20}. We suggest expanding the analysis of Up2Dep by incorporating more techniques such as static taint analysis (i.e., to identify sensitive data, track data flows, detect privacy-violating API calls), dynamic taint analysis (i.e., to monitor the application's behavior at runtime), and privacy policy compliance (i.e., to check for compliance with GDPR and CCPA, generate privacy reports). These add-ons may make Up2Dep a more robust tool for improving app security and privacy.

Up2Dep provides developers with information about vulnerabilities detected in third-party libraries using a manually updated database \cite{nguyen20}, which may become outdated or incomplete. On the other hand, the current version of DataAvalanche.io  does not contain a reliable database of alternative libraries and SDKs \cite{ekambaranathan23}, making it less practical for developers who may seek privacy-centered replacements. To address this, we propose creating a public, developer-maintained repository where developers can contribute vulnerability reports they encounter during development and recommend alternative libraries. For example, if a developer encounters a security or privacy vulnerability in a library like Solid.js \cite{solidjs2025}, they can submit a report to the repository, ensuring that other developers are properly informed, and they can suggest an alternative library such as React.js \cite{react2024}. This approach may help to keep the database more updated and comprehensive.

Additionally, even though DataAvalanche.io provides alternative libraries and SDKs, it does not provide a standardized framework to select the best-suited one for the application based on privacy implications. Therefore, we suggest creating a standardized framework for evaluating the privacy implications of SDKs and libraries, including criteria such as data-sharing practices, compliance with regulations like GDPR, and user reviews. It may allow developers to systematically compare and contrast different SDKs and libraries and make informed choices. 

Furthermore, since neither Up2Dep nor DataAvalanche.io has an integration with IDEs \cite{nguyen20}, it may require developers to manually check vulnerabilities, their privacy implications, and alternative library versions. Therefore, incorporating them with popular integrated development environments (IDEs) will streamline the development process and encourage developers to use these tools more regularly. For example, providing them with real-time feedback within the IDE, such as highlighting insecure dependencies as developers write code, may significantly enhance usability. For instance, if a developer includes a vulnerable version of a library, the IDE can immediately flag it and suggest a secure alternative. 

Lastly, as the authors mentioned, both solutions were evaluated on a small scale \cite{ekambaranathan23, nguyen20}, leaving concerns about their effectiveness across diverse applications. Therefore, more evaluations are needed in different industry conditions. For example, larger and more diverse groups of developers from different industries (e.g., healthcare, finance, etc.) and regions (e.g., Europe, Australia, etc.) may help identify potential gaps and ensure that the tools are flexible enough to meet the needs of a broader audience. 

\subsection{Make Software Privacy Compliant and Transparent}
\label{subsec:dis-privacy-statements}

As shown in Figure \ref{results}, we identified tools from three different categories that help developers make software privacy-compliant and transparent. Below, we discuss the implications of the results and potential future directions based on their limitations.

\subsubsection{Privacy label generation}
\label{subsubsec:dis-privacy-label}

We discussed in Section \ref{subsubsec:privacy-label-generation} how Privacy Label Wiz \cite{gardner22}, Matcha \cite{tianshi24}, and Jain et al.'s methodology \cite{jain23} helped developers in the privacy label generation process. Some of these solutions utilize static code analysis \cite{gardner22, tianshi24}. However, accurately capturing all data-handling practices needs some further implementation. The authors of both Privacy Label Wiz and Matcha believe that incorporating dynamic analysis techniques to observe runtime behavior may improve their accuracy since static analysis alone cannot capture all data flows or interactions \cite{gardner22, tianshi24}. For example, static analysis can identify that the application accesses the user's location, but it cannot determine whether the application actually collects location data at runtime, and the frequency of doing it. On the other hand, Dynamic analysis can monitor the application's behavior in real time. It means it can detect when the application accesses location data when the user interacts with a specific feature, such as a map. This may help developers provide more precise information about the privacy labels. Additionally, the authors of Matcha suggested incorporating a machine learning model, Codex (a large programming model), to enhance the analysis process \cite{tianshi24}. ML-driven dynamic code analysis and pattern recognition may help both tools identify privacy-sensitive code segments and track data flows more effectively, enhancing the dynamic analysis process. 

Further, PLW cannot determine what specific data third-party libraries collect that are integrated into an application \cite{gardner22}. Since privacy violations may occur at the library level, such as third-party libraries collecting or sharing user data without the developer's awareness \cite{nguyen20, ekambaranathan23}, this limitation may impact the reliability of generated privacy labels. For instance, Firebase might collect user analytics or device identifiers, but the static analyzer of PLW may not be able to identify these practices. Therefore, we suggest using synthetic data generated based on the application context and for different user interactions. For instance, a user logs into the app with synthetic data like email and password. Injecting this data into third-party libraries may help to simulate how libraries behave for different user interactions and identify what specific data the libraries collect. Furthermore, to improve the accuracy of detecting third-party libraries and data collection practices, PLW could be extended to include a centralized repository of privacy-related information (e.g., detailed documentation, privacy policies, and data collection practices provided by third-party library developers) for commonly used third-party libraries \cite{gardner22}. 

Finally, merging both Matcha and Privacy Label Wiz to make a combined solution for both Apple and Google platforms may help developers generate privacy labels for both platforms within a single solution.

On the other hand, Jain et al.'s fine-grained localization methodology posed some limitations \cite{jain23}, questioning its usability in practical software development scenarios. One of the main concerns is its limited evaluation in industry settings \cite{jain23}. Usually, developers have limited time to complete the software development project \cite{miryung04}. Since the proposed methodology is time-consuming and requires manual developer intervention \cite{jain23}, it may be effective to evaluate this in different industry settings where developers have short-term and long-term deadlines. In addition, since the evaluations were done for Android applications \cite{jain23}, it may not be effective for developers in organizations where they have the application for both Android and iOS versions, because they require another approach for iOS. Therefore, evaluating it on other platforms and ensuring its effectiveness may improve its usability. This also applies to Matcha and the Privacy Label Wiz.

\subsubsection{Privacy policy generation}
\label{subsubsec:dis-privacy-policy}
Under the privacy policy generation, we identified tools, PrivacyFlash Pro \cite{zimmeck2021PrivacyFlashPA} and Coconut \cite{coconut}, that especially target individual developers or small organizations. The major difference between these tools is their focus. For instance, Coconut helps developers understand and enhance their app's data-handling practices by providing real-time feedback during development, which helps them write better privacy policy statements. However, it does not automatically generate these statements \cite{coconut}. In contrast, PrivacyFlash Pro is designed to automate the entire privacy policy generation process \cite{zimmeck2021PrivacyFlashPA}. However, both solutions have some limitations that hinder their full effectiveness. 

As we discussed, PrivacyFlash Pro combines static code analysis with a questionnaire wizard \cite{zimmeck2021PrivacyFlashPA}. It implies that it relies on developer-provided answers. However, reliance on developers' responses may provide inaccuracies that may result in misleading privacy statements because the responses may depend on the developers' privacy knowledge. Since Coconut can provide real-time feedback and suggestions to improve data handling practices \cite{coconut}, merging that functionality of Coconut with PrivacyFlash Pro may increase the accuracy of developers' responses and ultimately make accurate privacy policies. We suggest improving Coconut's feedback mechanism by implementing more advanced analysis techniques, such as machine learning-based analysis, to give developers feedback based on their privacy knowledge by studying their development behavior. Also, providing feedback for any misalignment with regulatory frameworks like GDPR and CCPA may offer more specific guidance. At the same time, both tools mentioned that the code analysis process requires further improvement \cite{zimmeck2021PrivacyFlashPA, coconut}. The Python-based code analysis of PrivacyFlash Pro is not entirely accurate and produces some false positives and negatives \cite{zimmeck2021PrivacyFlashPA}. Also, developer-based evaluation results of Coconut showed that developers have a lack of trust in automatically filled values of annotations \cite{coconut}. These imply that the code analysis techniques are not accurate \cite{zimmeck2021PrivacyFlashPA, coconut}. Therefore, as we discussed in Sections \ref{subsubsec:dis-privacy-label}, \ref{subsec:dis-third-party}, and \ref{guideline-based-design-development}, we suggest trying ML-based approaches for the code analysis. One potential solution to address PrivacyFlash Pro's limitation of not considering server-side data sharing might be the integration of machine learning (ML) approaches as part of a dynamic analysis. For example, an ML model can be trained to analyze server logs and API calls in real time. Further, extending the PrivacyFlash Pro's platform support may increase the usability as we discussed for some other approaches in Sections \ref{sec:subsub:seccode} and \ref{subsubsec:dis-privacy-label}. Lastly, further evaluations are needed in different industry settings, as we discussed in Sections \ref{subsubsec:dis-privacy-label} and \ref{subsec:dis-third-party} to ensure wide usability.

\subsubsection{Privacy notice generation}
\label{subsubsec:dis-privacy-notice}
We discussed in Section \ref{subsubsec:privacy-notice-generation} a tool, Honeysuckle \cite{tianshi21}, that helps developers generate privacy notices for Android applications. Honeysuckle relies on developers to get accurate annotations \cite{tianshi21}, which could lead to misleading notices if developers do not provide accurate information. Additionally, Honeysuckle has limitations in its code analysis capabilities, such as being simple and adapted from the Coconut tool, and also needs developers' manual intervention \cite{coconut, tianshi21}, implying it may fail to detect complex data flows and third-party library interactions. For instance, if the application collects user data in one function, processes it in another, and then sends it to a third-party library, Honeysuckle may miss this data flow, leading to incomplete privacy notices. Developers' manual intervention may lead to errors, as we discussed in Section \ref{subsubsec:dis-privacy-policy}. At the same time, Honeysuckle provides limited support for third-party libraries (i.e., it might not fully capture the data-handling practice unless the library is explicitly supported) \cite{tianshi21}, reducing its effectiveness in real-world applications. Therefore, automating some aspects of annotation using code analysis techniques used by Matcha \cite{tianshi24}, PLW \cite{gardner22}, and PrivacyFlash Pro \cite{zimmeck2021PrivacyFlashPA} may provide better detection of privacy-related data flows, especially in libraries, and reduce developer workload while improving the accuracy. Additionally, we suggest using ML-driven analysis techniques as we suggested for other approaches in Sections such as \ref{subsubsec:dis-privacy-label}, \ref{subsec:dis-third-party}, \ref{guideline-based-design-development}, not only to analyze the code but also to generate privacy notice UIs based on the application domain and context. For example, notices (i.e., UI-based notices) for children's applications may differ from those for other applications (e.g., they should be more explainable) to provide them with more contextualized privacy information. Multiple language support may also be useful to help users make more informed decisions using their preferred language. Further, extending the support of Honeycucke for other web and mobile platforms may improve its usability, as we discussed in other Sections (e.g., Sections \ref{sec:subsub:seccode} and \ref{subsubsec:dis-privacy-label}).

\subsection{Improve Privacy Education and Awareness}
\label{subsec:dis-privacy-education}

We discussed how Privacy Ideation Cards (PIC) \cite{LugerEwa15, TangYing21} and workshop-based interventions \cite{weirCharles23} tried to improve the privacy education of software developers in section \ref{subsec:privacy-education}. However, their effectiveness across different developer experience levels and industrial settings is questionable because of their limitations. 

One of the limitations of PIC is that it has only been tested among software students \cite{TangYing21}. Even though the study showed that the students engaged in deeper discussions while using PICs \cite{LugerEwa15, TangYing21}, it is still questionable whether industry professionals who face tight deadlines and cognitive tasks \cite{miryung04, yasemin16, PedroBarbosa} would get the benefit of this in industry settings. Similarly, the workshop-based intervention proposed by Weir et al. \cite{weirCharles23} helped privacy-related communication between developers and product managers who have limited privacy expertise, raising concerns about its applicability among professionals who have some level of privacy expertise. Furthermore, these workshops' effectiveness relies on a supportive leader \cite{weirCharles23}. It may limit the applicability of this tool among organizations, as not all organizations may have a privacy-oriented leader. Here, the leader in this context refers to the development team's leader (e.g., architect, principal architect, or chief technology officer). Therefore, both solutions should be evaluated further in different industry settings, as discussed in Sections \ref{subsubsec:dis-privacy-label} and \ref{subsec:dis-third-party}.

Increasing privacy awareness in an organization may be challenging since a team will represent members with different mentalities. For instance, one may have strong privacy attitudes while one may have an "I have nothing to hide" mentality or negative privacy attitudes \cite{solove2007nothing, TahaeiMohammad21}. If they consider others' privacy in the same direction as they thought, it will affect software as well as the whole team's mentality \cite{TahaeiMohammad21}. Therefore, organizations should focus on educating developers about privacy attitudes, knowledge about human rights, social benefits, and empathy toward users to promote privacy in organizations (i.e., increasing privacy culture) \cite{TahaeiMohammad21}. 

In addition to that, it is recommended to educate developers about the practical implementation of privacy, targeting their roles in the development team rather than providing them general privacy awareness \cite{TahaeiMohammad21}. Design and code review, as well as mentoring programs, are the recommended approaches to educate practical privacy implementation \cite{TahaeiMohammad21}.

With these, we can argue that embedding privacy education into the software development process rather than taking it as an external training activity may help organizations develop a strong privacy culture.

\section{Threats to Validity} \label{theats_val}

Following the SLR guidelines proposed by Kitchenham and Charters, we identified several potential threats that could affect the validity of this review. Even though we made efforts to conduct the study rigorously, several risks may remain that readers should consider when interpreting the results.

\subsubsection{Selection and Inclusion Bias}

Although we used clearly defined inclusion and exclusion criteria, there is a possibility that some relevant studies were unintentionally excluded. For example, there may be some studies that addressed developer-supporting privacy solutions but using a different terminology (e.g., keywords) other than what we used in our search strategy. Additionally, the SLR is exposed to a publication bias as we focus solely on peer-reviewed academic sources. This could result in overlooking valuable insights from industry or non-academic sources.

\subsubsection{Coder Bias}

Although we followed a well-structured protocol (i.e., reflexive thematic analysis) to reduce the subjectiveness in decisions, the study may have introduced a coder bias when classifying the results, since there is a degree of human judgment in the process.

\section{Limitations} \label{limitations}

This SLR was conducted to offer a comprehensive overview of the topic while ensuring the reproducibility of the reported results in the literature. First, the returned articles for the search query were filtered using their titles and abstracts according to our inclusion and exclusion criteria. In that study selection phase, relevant articles may be overlooked. Therefore, to avoid such scenarios as much as possible, we conducted both forward and backward snowballing searches \cite{snowball14} for both articles selected from the results of the search query and from top-tier journals and conferences. Once the articles were ready, we thematically analyzed the selected articles to answer the research questions mentioned in Section \ref{introduction}. The first author performed the initial coding process and theme identification. However, the generated codes and themes may be influenced by the coder's experience, knowledge, and perspective, which potentially introduces bias. To minimize this bias, as recommended by Braun and Clarke \cite{Clarke2014}, we incorporated the perspectives of all authors when developing themes, as discussed in Section 3.3. Further, the SLR primarily focused on studies published in academic sources, which may result in the potential overlooking of contributions from industry practice. 

\section{Conclusion and future work} \label{conclusion}

This SLR explored and analyzed existing solutions, including tools, guidelines, methods, methodologies, and frameworks in the current literature to address the challenges that developers face in integrating privacy into software development while supporting them in the privacy integration process. Through this extensive review, we identified that the existing developer-supporting solutions have been proposed aiming to address a primary set of common developer challenges such as lack of privacy expertise among developers \cite{gardner22, tianshi24, zimmeck2021PrivacyFlashPA, coconut, tianshi21, AljeraisyAtheer24}, difficulties in translating privacy principles into technical implementations \cite{AlhirabiNada23, BaldassarreMaria, krstic24, PeixotoMarianaPCMTool, weirCharles23, TangYing21}, inadequate regulatory guidance \cite{AljeraisyAtheer24, PeixotoMarianaPCMTool, weirCharles23, TangYing21}, and the increasing privacy and security risks associated with third-party libraries and SDKs \cite{nguyen20, ekambaranathan23, GiffinDaniel12}. In Section \ref{subsec:privacy-requirement}, we discussed tools and methods to support developers in embedding privacy in the early stage of software development. It included tools like PCM-tool \cite{PeixotoMarianaPCMTool}, RMCM \cite{MAI2018165}, LINDUNN-Go \cite{WuytsKim20}, and a method called ThreatPoker \cite{RyggeHanne}. For design and development, tools like Parrot \cite{AlhirabiNada24} and Canella \cite{AljeraisyAtheer24} provide interactive environments for embedding privacy into IoT applications, while POSD (Privacy-Oriented Software Development) \cite{BaldassarreMaria} offers structured guidelines to facilitate privacy integration. Additionally, methodologies and frameworks have also been proposed to help developers in the software design and development process. It included frameworks such as CIA-level driven SDLC \cite{KangSooyoung}, secD4CloudMobile \cite{ChimucoFrancisco}, and Hails \cite{GiffinDaniel12}, and methodologies such as PbE \cite{PedroBarbosa}, CryptSDLC \cite{LoruenserThomas18}, and UML-based MDD \cite{krstic24}. To enforce privacy in coding, tools like FixDroid \cite{nguyen17} and PrivacyCAT \cite{privacycat} help detect vulnerabilities and suggest fixes to improve privacy in Android applications, and specifically in the WhatsApp application. Managing third-party dependencies is supported by tools like Up2Dep \cite{nguyen20} and DataAvalanche.io \cite{ekambaranathan23}, which help developers avoid insecure or non-compliant SDKs and libraries. Further, PrivacyStreams \cite{privacystreams} and Platys \cite{MurukannaiahPradeep} are tools and frameworks, respectively, that have been proposed to help developers manage and secure personal data when developing software applications. Furthermore, to assist developers in understanding privacy principles, regulations, and secure development practices, approaches like Privacy Ideation Cards (PICs) \cite{TangYing21} and a workshop-based \cite{weirCharles23} intervention have been proposed. Finally, generating privacy statements and policies is made easier with tools like Matcha \cite{tianshi24}, Privacy Label Wiz \cite{gardner22}, PrivacyFlash Pro \cite{zimmeck2021PrivacyFlashPA}, Coconut \cite{coconut}, and Honeysuckle \cite{tianshi21} which assist developers in generating accurate privacy labels, policies, and notices to align with compliance requirements.

However, the identified tools, guidelines, methods, methodologies, and frameworks discussed in the results section (i.e., Section \ref{results}) offer only partial solutions because of the limitations we discussed in the discussion section (i.e., Section \ref{discussion}). Many solutions require prior expertise in privacy and security \cite{MAI2018165, PedroBarbosa, coconut}, and require developers' manual intervention \cite{PeixotoMarianaPCMTool, jain23, BaldassarreMaria, zimmeck2021PrivacyFlashPA, coconut, tianshi21, KangSooyoung}, posing a barrier for developers with limited knowledge in these areas to use these solutions, as well as the possibility of human errors. Some solutions do not provide seamless integration with agile and real-world development workflows \cite{PeixotoMarianaPCMTool, BaldassarreMaria}, which causes difficulties in adoption for rapid software development environments. Additionally, many solutions are designed for specific environments (e.g., Android) or programming languages (e.g., Java) \cite{privacystreams, BaldassarreMaria, krstic24, GiffinDaniel12, nguyen17, tianshi24, gardner22, zimmeck2021PrivacyFlashPA}, limiting their applicability across diverse software development ecosystems. Further, most of the solutions need further evaluations in different industry settings \cite{PeixotoMarianaPCMTool, WuytsKim20, RyggeHanne, nguyen20, ekambaranathan23, AlhirabiNada24, BaldassarreMaria, PERERA2020238, LoruenserThomas18, ChimucoFrancisco, privacycat, tianshi24, gardner22, jain23, TangYing21}. Furthermore, some of the solutions provide their support for a selected set of regulations or do not cover the entire regulation, limiting their usability \cite{AljeraisyAtheer24, AlhirabiNada24}. We discussed in detail how these limitations affect the developers in Section \ref{discussion}. These limitations imply that they fail to address the developer challenges and necessitate that additional research is required. 

Therefore, as discussed in the discussion section, we propose future improvements to each developer-supporting approach to make them more effective. First, since we identified that developers' lack of privacy knowledge and inaccuracy of the analysis mechanisms (e.g., code analysis) as critical concerns in the proposed solutions, we suggested to incorporate machine learning approaches (e.g., NLP), to reduce the developers' manual intervention by automating the integrated processes, increase the accuracy (e.g., the accuracy of privacy label generation), and also to reduce the human errors happen because of the lack of privacy knowledge. However, because machine learning solutions are not 100\% accurate, developers' feedback may be important at some point. This implies that developers' privacy knowledge may be critical at some point. Therefore, future research should be more focused on enhancing developers' privacy knowledge either as a direct or an indirect goal. Here, indirect refers to the fact that the privacy knowledge is improved while developers engage in some other task, such as privacy policy generation. We discussed how we can provide this for some of the existing solutions in detail in the discussion Section (i.e., Section \ref{discussion}. Next, we suggested expanding some solutions in a way to support more programming languages (e.g., Dart, C++) and platforms (i.e., Android, iOS) in order to increase usability. It helps developers and organizations who work with different tech stacks to rely on the same tool, guideline, method, methodology, or framework regardless of the tech stack. Additionally, we suggested extending some of these solutions to support broader regulatory compliance (e.g., GDPR - Europe, CCPA - USA, etc.) as it is essential for developers and organizations who develop software applications targeting different regions. Further, we suggested conducting extensive empirical evaluations for most of the existing solutions to check their effectiveness and suitability in different industry settings. We explained in detail what kind of evaluations are needed in Section \ref{discussion}.

Finally, regarding the SLR itself, we acknowledge that our study mainly focused on peer-reviewed academic literature. To further improve the comprehensiveness of future SLRs, we suggest incorporating grey literature sources such as technical blogs, industry reports, and open-source repositories. Including these sources will help to explore popular but non-academic solutions (e.g., tools, methods, etc) which offer additional insights about real-world privacy integration challenges and solutions.

In summary, while existing developer-supporting solutions provide developers with assistance in integrating privacy into software development, their limitations highlight that further enhancements are needed. Therefore, addressing these gaps, as we discussed, will be crucial to making these solutions more effective and accessible for developers across diverse software development environments.

\bibliographystyle{ACM-Reference-Format}
\bibliography{sample-base}

\end{document}